# Generalized Brewster-Kerker effect in dielectric metasurfaces

*Ramón Paniagua-Domínguez[1]\*†, Ye Feng Yu[1]†, Andrey E. Miroschnichenko[2], Leonid A. Krivitsky[1], Yuan Hsing Fu[1], Vytautas Valuckas[1,3], Leonard Gonzaga[1], Yeow Teck Toh[1], Anthony Yew Seng Kay[1], Boris Luk'yanchuk[1], and Arseniy I. Kuznetsov[1]‡*

[1]Data Storage Institute, A\*STAR (Agency for Science, Technology and Research),
5 Engineering Drive 1, 117608, Singapore

(corresponding authors: \*ramon-paniag@dsi.a-star.edu.sg; ‡arseniy_k@dsi.a-star.edu.sg)

[2]Nonlinear Physics Centre, Research School of Science and Engineering, The Australian National University, Acton, ACT, 2601, Australia

[3] Department of Electrical and Computer Engineering, National University of Singapore,
1 Engineering Drive 2, 117576, Singapore

†These authors contributed equally to this work.

**Polarization is one of the key properties defining the state of light. It was discovered in the early 19th century by Brewster, among others, while studying light reflected from materials at different angles. These studies led to the first polarizers, based on Brewster's effect. One of the most active trends in photonics now is devoted to the study of miniaturized, sub-wavelength devices exhibiting similar, or even improved, functionalities compared to those achieved with bulk optical elements. In the present work, it is theoretically predicted that a properly designed all-dielectric metasurface exhibits a generalized Brewster's effect potentially for any angle, wavelength and polarization of choice. The effect is experimentally demonstrated for an array of silicon nanodisks at visible wavelengths. The underlying physics of this effect can be understood in terms of the suppressed scattering at certain angles that results from the interference between the electric and magnetic dipole resonances excited in the nanoparticles, predicted by Kerker in early 80s. This reveals deep connection of Kerker's and Brewster's legacies and opens doors for Brewster phenomenon to new applications in photonics, which are not bonded to a specific polarization or angle of incidence.**

The oldest, and probably simplest, way to obtain linearly polarized light starting from unpolarized one is impinging it on a dielectric interface at the so called Brewster's angle. In this way, the reflected light will only have electric field component parallel to the interface. Well understood since the 1820's after the pioneering work of Fresnel, and experimentally known since the early 1810's from works of Malus and Brewster[1] (see also Ref. 2 for a succinct historical perspective), the Brewster's angle for homogeneous isotropic non-magnetic media can be defined as the angle for which Fresnel's reflection coefficient for *p*-polarized light (i.e., with the electric field parallel to the plane of incidence) vanishes, $\mathfrak{R}_p = 0$. An alternative definition states that Brewster's angle is the one at which the reflected and refracted waves are orthogonal, thus fulfilling the condition $\theta_i + \theta_t = \pi/2$, where $\theta_i$ is the angle of incidence and $\theta_t$ is the angle of refraction/transmission. The common microscopic interpretation of this effect is illustrated in Fig.1a. The induced electric dipoles, generated inside the medium in response to the driving electromagnetic wave, oscillate along the direction of the electric field perpendicular to the propagation direction. As the far field power radiated by a dipole vanishes along its oscillation axis, whenever the dipole and reflection direction are parallel, no radiation is emitted into that direction and reflection is inhibited. In all other directions apart from that of refraction, radiation is compensated by the rest of the dipoles within the medium. If polarization is switched, as shown in Fig.1b, due to the non-zero radiation in the plane perpendicular to the dipole, it is clear that such effect cannot be achieved.

The situation becomes more interesting when one considers a material which has both electric and magnetic dipoles excited in response to the electric and magnetic components of the incident wave. Such materials should have both electric permittivity and magnetic permeability different from unity ($\varepsilon \neq 1$, $\mu \neq 1$). In this case, the radiation pattern is no longer zero in the direction of oscillation of any of the orthogonal electric or magnetic dipoles (due to the non-zero contribution of the orthogonal dipole) and thus the classical Brewster effect can no longer be observed. Instead, there can be other particular directions at which the collective radiation of both dipoles vanishes due to their destructive interference, as predicted by Kerker and co-authors in early 80s[3]. These directions are determined by the relative amplitudes and phases of the dipoles. In the macroscopic picture, this interference may lead to the appearance of an analogue to Brewster's angle defined by both electric and magnetic properties of the material, as depicted in Fig.1c. This is, the ratio $\varepsilon/\mu$ determines the angle at which the condition $\Re_{s,p} = 0$ is satisfied[4]. More importantly, inhibition of radiation from a pair of dipoles can happen at any angle and in any of the two oscillation planes depending on their relative amplitudes and phases. Thus, for such a material Brewster's angle may exist, potentially, for any of the two polarizations and at any angle of incidence (even below 45 degrees without leading to total internal reflection at some higher angles). Both polarizations cannot, however, simultaneously have zero reflection for a given angle, except for the very particular case of $\varepsilon = \mu$ (impedance matched) at normal incidence[5]. In this case, each polarizable portion of matter will have induced electric and magnetic dipoles having the same amplitude and phase leading to inhibition of backscattered radiation, i.e., fulfilling the so called first Kerker's condition, originally derived for small magnetic particles[3]. In case of purely magnetic media, $\mu \neq 1$ and $\varepsilon \approx 1$, one can find a situation when the analogue to Brewster's angle appears for s-polarized light, having the magnetic field vector parallel to the plane of incidence, which is orthogonal to the conventional Brewster effect in dielectric media. Some of the important phenomenology associated to the generalized Brewster effect can be found in the Supplementary Information, Section 1.

All this findings remained a mere theoretical curiosity for almost 20 years, since for natural materials the magnetic response is typically very weak at optical frequencies ($\mu \approx 1$). Nevertheless, since the advent of metamaterials new ways to produce optical magnetic response have been explored[6-8]. As a result some attempts have been done towards finding Brewster's angle in s-polarization in bulk magnetic metamaterials, both in microwaves[9] in arrays of split ring resonators and at optical frequencies in strongly anisotropic media[10]. Recently, polarization rotation in reflection from meta-films of bi-anisotropic split rings has been theoretically studied at microwave frequencies in connection to Brewster effect[11].

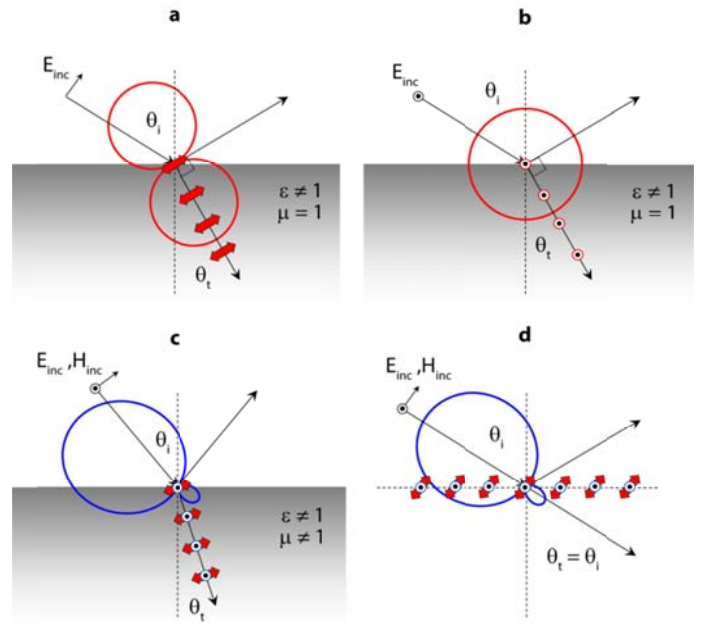

Figure 1. Microscopic interpretation of Brewster effect and proposed metasurface. a, p-polarized light impinging on a dielectric medium, $\varepsilon \neq 1$ and $\mu = 1$, under usual Brewster's condition. Red line shows 2D emission diagram of electric dipoles excited inside the material by the refracted wave. b, Same as in a but for s-polarized incidence, for which no Brewster effect can be observed. c, Generalized Brewster effect for a dielectric medium with electric and magnetic response, $\varepsilon \neq 1$ and $\mu \neq 1$. Blue line shows 2D emission diagram of interfering electric and magnetic dipoles excited inside the material by the refracted wave. d, Generalized Brewster-Kerker effect in the proposed array of high refractive index nanoparticles.

In this paper, it is demonstrated both theoretically and experimentally that the generalized Brewster effect can be observed, potentially, in any two-dimensional sub-diffractive arrangement of high-index dielectric nanoparticles, or in any other system where strong electric and magnetic resonances can be efficiently excited. It is shown that this effect is a direct consequence of the angle-suppressed radiation/scattering due to interference between the electric and magnetic dipoles excited in the particles within the array (see schematic in Fig.1d), thus connecting two apparently unrelated phenomena such as Brewster's angle and Kerker's conditions. We thus call it Brewster-Kerker effect. Silicon (Si) nanoparticles are specifically considered, for which such resonances have been broadly studied both theoretically[12-14] and experimentally[15-17]. They have attracted particular attention within the field of artificial magnetism at optical frequencies[12-17] due to their low intrinsic losses and CMOS compatibility which holds promise for finding real world applications. Their exciting properties regarding magnetic near-field enhancement[18-22] and directional scattering[23-26], together with their low dissipation, makes them ideal nanoantennas for visible and near-infrared light[27]. The possibility to realize the first Kerker's condition[23-25] has also inspired studies on using them as ideal Huygens' sources in highly-efficient transmissive metasurfaces[28-30]. Also their strong interaction with light, leading to high reflection and phase accumulation, makes them ideal candidates to act as efficient reflectors or phase-controlled mirrors[12,31-33]. The present study comes to extend this already broad realm with new fascinating properties. Moreover, novel generalized Brewster phenomenon giving great degree of freedom in polarization and incident angles may open doors to multiple new applications in photonics, which could not be achieved with standard Brewster effect in conventional dielectric media.

## 1. Generalized Brewster-Kerker effect in two-dimensional arrays of silicon spheres.

Let us start considering a single silicon nanosphere under plane wave illumination (see Figs.2a and 2b), for which the required electric and magnetic dipole modes can be efficiently excited. The scattering cross section ($C_{sca}$) for a sphere with diameter $D = 180$ nm, as computed analytically with Mie theory[34], is depicted in Fig.2c. Partial scattering cross-sections by the first excited resonant modes, namely the electric (ED) and magnetic (MD) dipoles and the electric (EQ) and magnetic (MQ) quadrupoles are also shown. As can be seen, the usual hierarchy of resonances in high-contrast dielectric nanoparticles starts with the lowest-energy magnetic dipole followed by the electric dipole mode[12-16]. Thus, whenever higher order modes are negligible each sphere can be accurately described by a pair of these dipoles.

Kerker and co-workers[3] showed that, in such systems, the scattered far-field can be completely polarized parallel or perpendicular to the scattering plane in some particular observation direction, and this direction depends on the relative strength of the induced electric ($\boldsymbol{p}$) and magnetic ($\boldsymbol{m}$) dipoles. Originally derived for magnetic spheres, this result relates to interference in the electric far-field radiated by a pair of such dipoles, which reads:

$$\boldsymbol{E}_{ff} = \boldsymbol{E}_{ff}{}^p + \boldsymbol{E}_{ff}{}^m = \frac{k_0^2}{4\pi\epsilon_0}\left[\hat{\boldsymbol{n}} \times (\boldsymbol{p} \times \hat{\boldsymbol{n}}) + \frac{1}{c}\boldsymbol{m} \times \hat{\boldsymbol{n}}\right] \quad (1)$$

with $k_0 = 2\pi/\lambda$ the wavenumber and $\epsilon_0$ and $c$ the permittivity and speed of light in vacuum, respectively, and $\hat{\boldsymbol{n}}$ the unit vector in the observation direction.

Consider now the particular situations depicted in Figs.2a and b. It also follows from (1), see section 2 in Supplementary Information, that in the plane containing the incident wave-vector and the induced electric dipole (highlighted in Fig.2a), the radiated electric field vanishes in the observation direction defined by the angle $\theta$ if:

$$\cos(\theta - \theta_i) = m/p \quad (2)$$

in which $p$ and $m$ are the complex amplitudes of electric and magnetic dipoles. In the orthogonal plane, which contains the incident wave-vector and the induced magnetic dipole (case depicted in Fig.2b), the field vanishes when:

$$\cos(\theta - \theta_i) = p/m \quad (3)$$

Note that the backscattering direction is defined by $\theta = \theta_i$. In this direction, the field vanishes when $p = m$ (first Kerker's condition[3]). Note also that equations (2) and (3) are, in general, complex and become real only when the dipoles are in phase or anti-phase. From these equations, it can be seen that radiation can be totally suppressed for angles in backward directions $|\theta - \theta_i| \leq \pi/2$ exclusively if the dipoles are in phase ($p$ and $m$ having the same sign) and in forward directions ($|\theta - \theta_i| \geq \pi/2$) if they are in anti-phase ($p$ and $m$ having opposite sign). The spectral regions in which the induced dipoles are approximately in phase or anti-phase for the silicon sphere are highlighted in Fig.2c by yellow and green shading colours, respectively. They indicate the spectral ranges for which scattering cancellation in forward and backward directions may happen.

The partial scattering cross sections[34] associated with the electric ($C_{sca}^{ED}$) and magnetic ($C_{sca}^{MD}$) dipoles are proportional to the squared modulus of the dipole moments ($C_{sca}^{ED} \propto |\boldsymbol{p}|^2$ and $C_{sca}^{MD} \propto |\boldsymbol{m}|^2$), and this allows to recast equations (2) and (3) as:

$$C_{sca}^{MD}/C_{sca}^{ED} = |\cos(\theta - \theta_i)|^2 \quad (4)$$

$$C_{sca}^{ED}/C_{sca}^{MD} = |\cos(\theta - \theta_i)|^2. \quad 5)$$

It immediately follows from (4) that the electric dipole scattering must dominate ($C_{sca}^{MD}/C_{sca}^{ED} < 1$) to achieve cancellation in the plane containing the electric dipole. Similarly, it follows from (5) that the magnetic dipole scattering should be dominant ($C_{sca}^{ED}/C_{sca}^{MD} < 1$) to achieve the scattering cancellation in the plane containing the magnetic dipole. The regions of dominant electric and magnetic dipoles are highlighted in Fig.2c by red and purple shading colours, respectively.

In Fig.2d the 2D scattering pattern of the Si sphere computed from Mie theory is plotted for two selected wavelengths, $\lambda_1 = 614$ nm and $\lambda_2 = 728$ nm, in the plane containing the incident wave-vector and the electric or magnetic dipole, respectively. Vanishing scattering intensity angles predicted by equations (4) and (5), respectively, are also shown. At $\lambda_1$ the ED dominates and the dipoles are in anti-phase leading to scattering cancellation at an angle $|\theta - \theta_i| \geq \pi/2$ in the plane containing the incident electric field.

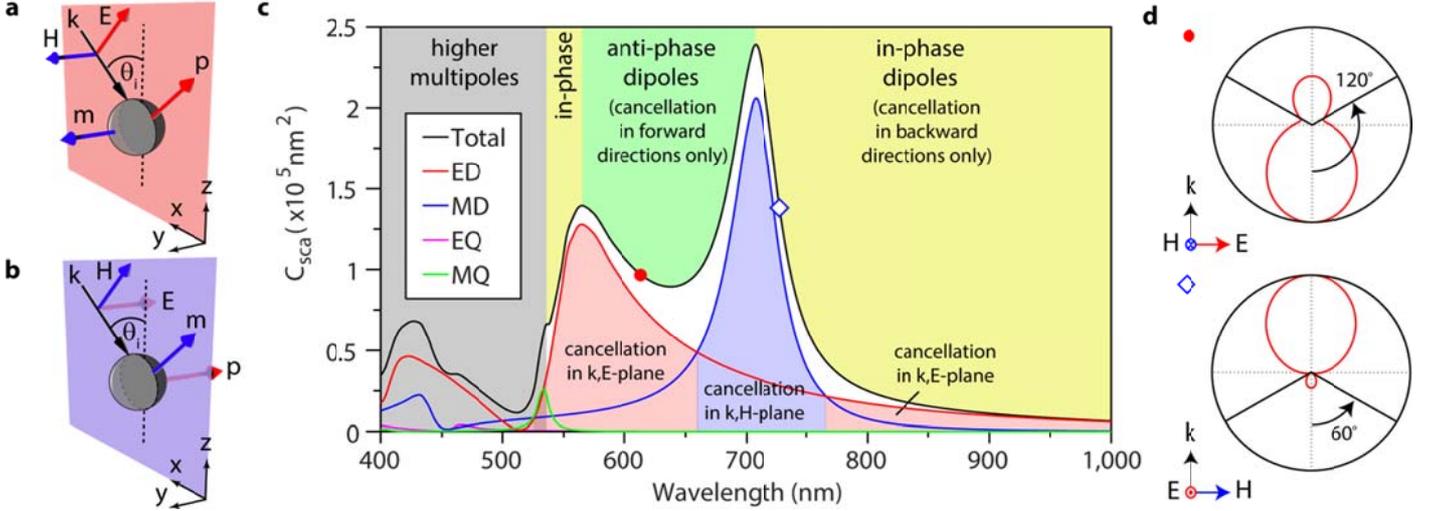

**Figure 2. Optical properties of a silicon spheres with diameter $D = 180$ nm in air under plane wave illumination.** **a-b.** Schematic representation of the two situations studied. **c,** Total scattering cross section (black curve) and contributions from the electric dipole (red curve), magnetic dipole (blue curve), electric quadrupole (magenta curve) and magnetic quadrupole (green curve). The different shaded regions indicate those frequency windows for which the induced electric and magnetic dipoles are approximately in phase (yellow) or in anti-phase (green), and those for which the electric dipole (light red) or the magnetic dipole (light blue) dominate over the other. **d,** Far-field radiation patterns for two wavelengths leading to inhibition of radiation at 60 degrees with respect to the forward- ($\lambda_1 = 614$ nm, red solid circle) and backward- ($\lambda_2 = 728$ nm, blue hollow diamond) scattering directions. In the first case zero radiation is only possible in the plane parallel to the electric field whereas in the second it is only possible in the perpendicular one.

At $\lambda_2$ the MD contribution is dominant and dipoles are in phase leading to scattering cancellation at an angle $|\theta - \theta_i| \leq \pi/2$ in the plane containing the incident magnetic field. Relative amplitudes and phases are such that interference suppresses radiation at 60 degrees with respect to the forward- and back-scattering directions, respectively.

Let us now consider the case of similar spheres arranged in an infinite two-dimensional sub-diffractive array in the *xy*-plane (see Fig.3a) under plane wave oblique incidence. It is clear that for *p*-polarized light the plane of incidence coincides with the plane that contains the incident electric field and the induced electric dipoles (as in Fig.2a). Correspondingly, for *s*-polarization the incidence plane contains the magnetic field and the induced magnetic dipoles (as in Fig.2b). Although the *effective* dipoles induced in the particles in the array are different from the single particle case due to the lattice interactions[35], they still radiate according to equations (1)-(3) in the plane of incidence. Note that interference from different sites in the infinite array makes radiation of the whole system allowed only as plane waves along the diffraction directions. In the case of sub-diffractive arrays, this implies radiation in the reflection and transmission directions only. Therefore, if the induced dipoles do not radiate along the direction of reflection, no reflection at all will occur in the system, leading to the Brewster's condition (see Supplementary Section 3 for a demonstration in the context of phased arrays). Following the discussion above, in such systems this may happen for both *s*- and *p*-polarized incident waves.

In the following, we consider an infinite square lattice of silicon spheres with diameter $D = 180$ nm and period $P = 300$ nm, as depicted in Fig.3a, and study its reflection properties as a function of the wavelength and angle of incidence. We start with *p*-polarized light. Simulated results obtained by means of Finite Element Method (see Methods section for details) are shown in Fig.3b. At normal incidence, the electric and magnetic resonances of the particles lead to the appearance of well-known bands of high reflectivity[12,31-32]. However, oblique incidence strongly changes this behaviour. At high angles of incidence one can observe three regions of extremely low reflection. The first one is a narrow region located at the blue side of the resonances (~515 nm). It is present at normal incidence and slightly redshifts for increasing angles (from 0 up to ~40 degrees). The second one, located in the red side of the resonances (~790 nm), is also present at normal incidence and strongly redshifts with increasing angles. Finally, a broad region, both in bandwidth (~150 nm) and angles of incidence (from ~40 to ~80 degrees), appears at higher angles, spectrally located between the positions of the electric and magnetic dipole resonances observed at normal incidence. Importantly, the angle of minimum reflection strongly depends on the wavelength and varies in the wide range. Figure 3c shows the angular dependence of reflection at some selected wavelengths to better illustrate this effect.

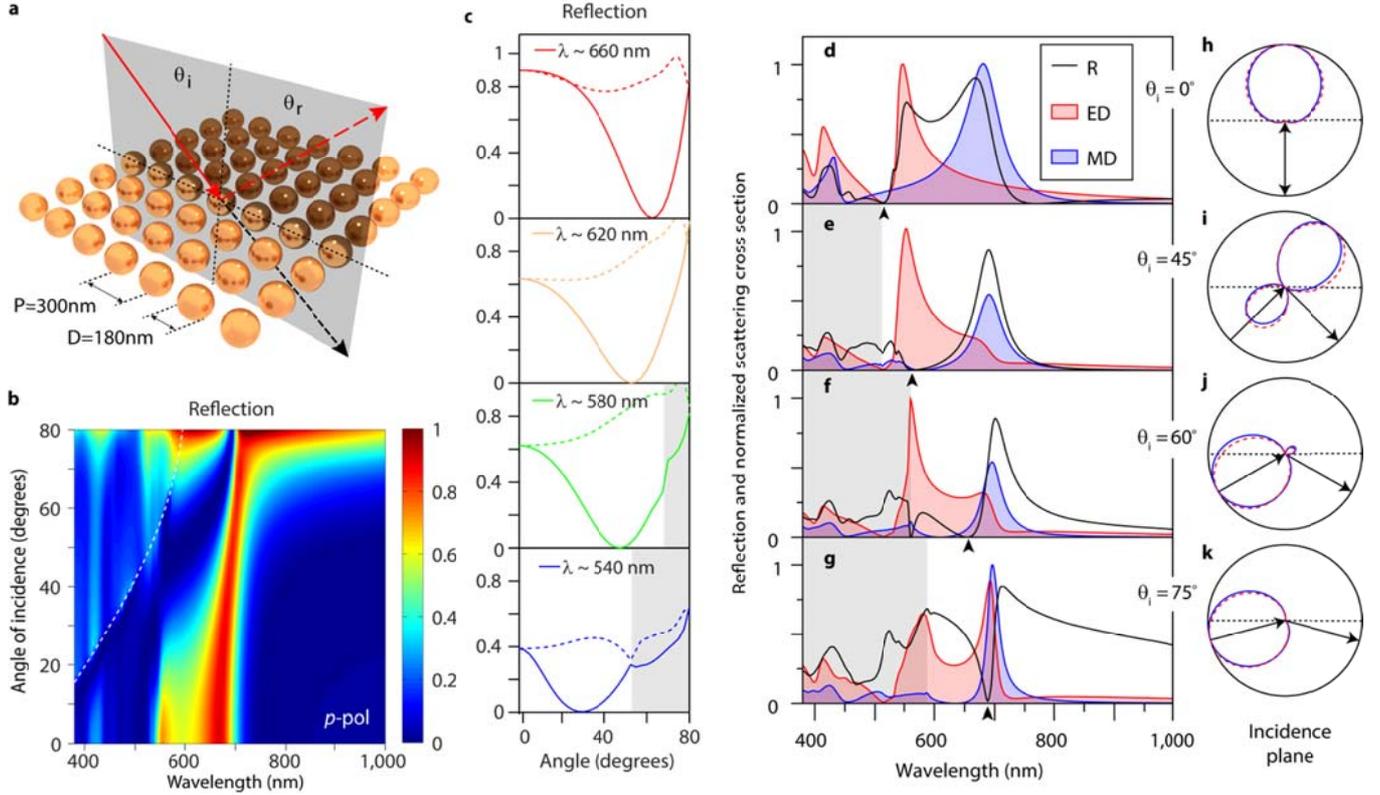

**Figure 3. Simulated optical response of a square lattice of silicon spheres under *p*-polarized oblique incidence. a,** Scheme of the simulated system. **b,** Numerically calculated reflection versus wavelength and angle of incidence for a square array of silicon spheres with diameter $D = 180$ nm and pitch $P = 300$ nm. **c,** Reflection versus angle of incidence for selected wavelengths showing the strong dependence of Brewster's angle on wavelength as well as the possibility of achieving values below 45 degrees. Solid lines correspond to *p*-polarization while dashed lines are the corresponding curves for *s*-polarization. Grey shaded areas mark the spectral regions affected by diffraction. **d-g,** Reflection (black curve), together with electric dipole (red curve and corresponding shaded area) and magnetic dipole (blue curve and corresponding shaded area) contributions to scattering (normalized to their common maximum) as a function of wavelength for the cases of normal incidence and oblique incidence with $\theta_i = 45, 60$ and 70 degrees respectively. Diffractive region is indicated by a shaded grey area. **i-l,** Radiation patterns in the plane of incidence numerically computed via Stratton-Chu formulas (blue solid curve) and from the induced electric and magnetic dipoles only (red dashed curve) at the wavelength of minimum reflection (arrowheads in d-g). Incidence and reflection direction are shown by arrows.

One can observe that reflection of *p*-polarized light (solid lines) turns into zero at some particular angle of incidence, resembling the conventional Brewster effect in dielectric media, while no special features are observed for *s*-polarized light at this angle (corresponding dashed lines). However, there are two major peculiarities of this system, which should be highlighted. First, the range of angles at which the reflection minimum is observed covers almost the whole 0-90 degrees span, not being restricted to angles above 45 degrees (opposite to the conventional Brewster effect). Second, as will be shown next, the effect is not restricted to *p*-polarization, thus gathering the main features of generalized Brewster phenomenon. It is important to mention that this effect is not related to diffraction. The first non-zero diffraction order, indicated as a dashed white line in Fig.3b and as shaded regions in Fig.3c, appears out of the range of wavelengths and angles for which the effect is observed.

Let us now focus on the spectral region between the electric and magnetic dipole resonances. In Figs.3d-g the case of normal incidence is shown, together with some cases with zero reflection in that region, namely 45, 60 and 75 degrees incidence. For normal incidence both reflection maxima spectrally coincide with the excitation of dipolar resonances inside the particles. Zero reflection is observed at 775 nm, where the induced electric and magnetic dipoles have the same amplitudes and phases meeting the first Kerker's condition[3,23-25], and at 515 nm, which is close to the Kerker's condition but also affected by higher-order contributions. In the cases of oblique incidence, zeros in reflection are observed at 566 nm for 45 degrees, 657 nm for 60 degrees, and 686 nm for 75 degrees (indicated by arrow-heads in Figs.3d-g) showing the strong wavelength dependence.

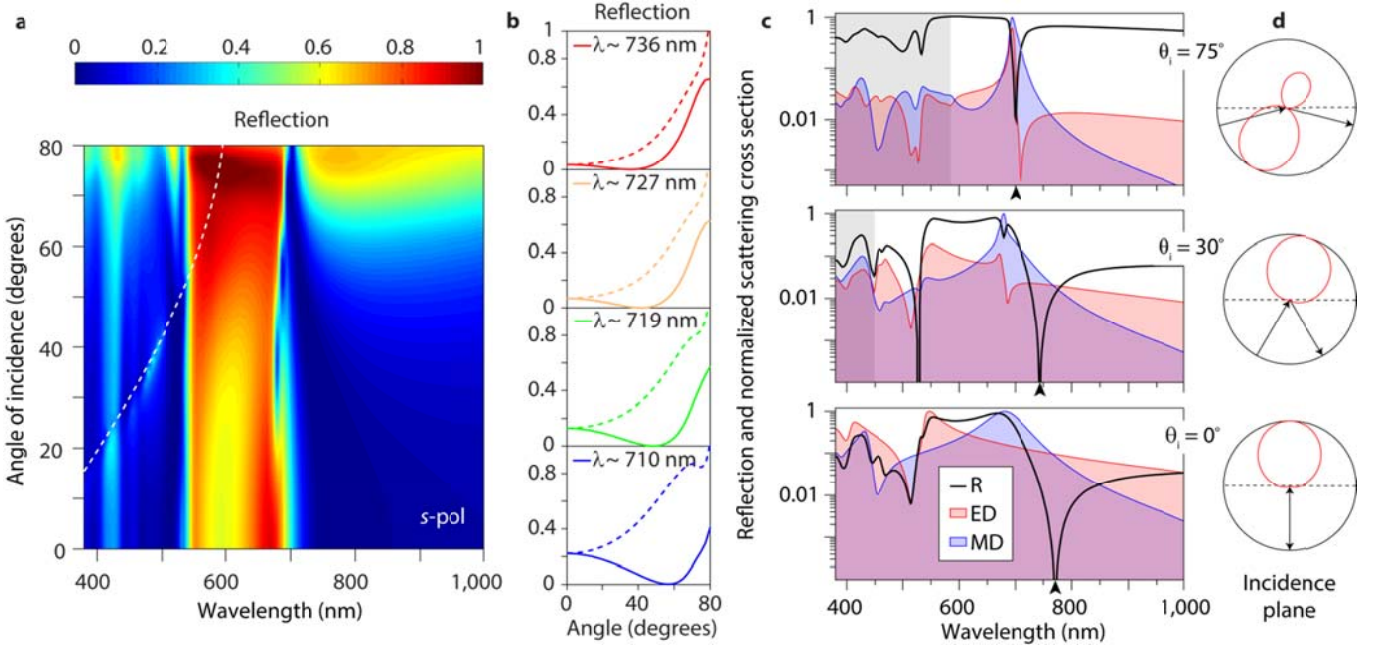

**Figure 4. Simulated optical response of a square lattice of silicon spheres under *s*-polarized oblique incidence. a,** Numerically calculated reflection versus wavelength and angle of incidence for a square array of silicon spheres with diameter $D = 180$ nm and pitch $P = 300$ nm. The wavelength for which the first diffracted order appears is indicated by the white dashed line. **b,** Reflection versus angle of incidence for selected wavelengths showing Brewster's angle for s-polarization. Solid lines represent s-polarization while dashed lines are the corresponding curves for p-polarization. Dependence on wavelength and the possibility to achieve values below 45 degrees are observed. **c,** Reflection (black curve) and electric dipole (red curve and corresponding shaded area, ED) and magnetic dipole (blue curve and corresponding shaded area, MD) contributions to the scattering (normalized to their common maximum) from a single sphere in the array for several angles of incidence. $\text{Log}_{10}$ scale is used for better visualization of the minima. Diffractive regions are indicated by the shaded grey area. **d,** Associated radiation patterns of each single particle in the array in the plane of incidence at the wavelengths of minimum reflection (arrowheads in **c**).

One explanation of the emergence of the reflection minimum at higher angles could be associated with disappearance of the dipolar resonances at off-normal incidence. However, this is not the case. Both dipolar modes are still efficiently excited, and it is their mutual interference which results in the radiation inhibition in the reflection direction. Figs.3d-g show by red and blue curves (and corresponding shaded areas), the ED and MD contributions to the total scattering cross-section ($C_{sca}$) from each single particle in the array, computed using the multipole decomposition technique, as explained in Methods. As readily seen, both dipole modes are present for those angles and wavelengths for which the reflection vanishes. The dipolar contributions are dominant and higher order modes only appear at shorter wavelengths (the complete map can be found in Section 4 in Supplementary Information). Interestingly, multipole decomposition reveals that the ED dominates in all the above cases. This is expected from equation (4) to be able to cancel radiation in the plane containing the electric field, which in *p*-polarized case coincides with the plain of incidence. In a simplified case with no interaction between the particles in the array, the induced dipoles should oscillate parallel to the incident field. In that case, in order to cancel scattering at the reflection angle $\theta = \theta_r = -\theta_i$, equation (4) imposes $C_{sca}^{MD}/C_{sca}^{ED} = \{0, 0.25, 0.75\}$ for $\theta_i = \{45, 60, 75\}$ degrees, respectively. Thus, usual Brewster at 45 degrees is covered in this description and requires vanishing of magnetic dipole for p-polarization, as expected. Actual values retrieved from simulations, in which interparticle interaction is taken into account, become $C_{sca}^{MD}/C_{sca}^{ED} = \{0.0078, 0.24, 0.63\}$, which are quite close to the interaction-free case. The second zero in reflection observed at 560 nm for $\theta_i = 60$ degrees is related to the onset of the diffractive regime (indicated as grey-shaded areas).

As a test of consistency, far-field radiation patterns from each single particle in the infinite array were computed using Stratton-Chu formulas[5] from the fields on the surface of the sphere and plotted in the plane of incidence in Figs.3h-k (blue solid lines). Also shown (red dashed lines) are the patterns radiated by the pair of electric ***p*** and magnetic ***m*** dipoles given by the multipole decomposition, computed through equation (1). Both patterns closely coincide and show zero radiation in the direction of the reflected wave (indicated, together with the incident one by arrows), thus confirming the dipole interference origin of the vanishing reflection regions.

Let us now switch to the case of *s*-polarized incidence to show that similar effects can be obtained. The change in polarization makes the plane of incidence coincide with that containing the magnetic field in the analysis for a single sphere, thus obeying equations (3) and (5). The simulated reflection versus wavelength and angle of incidence for the same array of spheres in *s*-polarized case is shown in Fig.4a. Two narrow-band frequency windows of vanishing reflection, shifting very weakly with the angle of incidence, can be observed starting at around 515 nm and 770 nm for normal incidence. Also, an omni-directional, high reflectivity region is observed in between, analogous to that reported for high index infinite cylinders[36]. Brewster effect in this polarization is evidenced by plotting, as in Fig.4b, the reflection against the angle of incidence for several wavelengths. We focus on the narrow band observed between 700 nm – 750 nm, for which no higher order multipoles are present. For *s*-polarized light (shown as solid lines) emergence of Brewster's angle is apparent, while no special features are observed for *p*-polarization (dashed lines). As readily observed, strong dependence on wavelength and span over the whole 0-80 degrees simulated range are also observed for *s*-polarization.

Now we show that the origin of Brewster's angle in this polarization is totally analogous to that of *p*-polarization. To this end, particular angles are plotted in Fig.4c together with the ED and MD partial scattering cross sections (normalized to their common maximum). For normal incidence spectral position of the dip corresponds to the first Kerker's condition at which electric and magnetic dipoles have similar amplitude and phases[3, 23-25]. The observed weak blue-shift of this dip with increased angle of incidence is a consequence of the particular shape of the resonances excited in the particles and their mutual interplay, which allows fulfilling equation (5) for every angle in a narrow spectral region. Note that within the whole range of wavelengths and angles with vanishing reflection, the MD contribution is higher than the ED one, as predicted by equation (5). Similar to the case of *p*-polarized incidence, the radiation patterns of each single particle in the array associated with zero-reflection wavelengths show no radiation in the reflection direction, thus confirming the interference origin of the effect also in *s*-polarization as depicted in Fig.4d.

It is important to stress that the observed spectral and angular behaviour of the zero reflection regions in the metasurface (Figs.3b and 4a) can be directly related to the scattering properties of the single building-blocks through amplitudes and phases of the induced dipoles, as described in detail in Section 5 of the Supplementary Information (and Fig.S7 therein). Thus, engineering these parameters, e.g. through the geometry of the inclusions, could lead to the generalized Brewster effect, potentially, at any desired angle, frequency and polarization of interest.

## 2. Experimental verification with arrays of silicon nanodisks.

To experimentally demonstrate the Brewster-Kerker effect, an array of silicon nanodisks was fabricated on a fused silica substrate (as described in Methods) through silicon film deposition, electron beam lithography and etching. Disks are chosen for ease of fabrication and, for aspect ratios close to unity, they are expected to have similar optical properties to spheres. The actual diameter is around $D = 180$ nm, height $H = 150$ nm and array pitch $P = 300$ nm (see SEM images of the fabricated array in the insets to Fig.5a). Angular-dependent reflection measurements were performed using a home built free-space microscopy setup (see Methods for details). The measured reflection and transmission spectra under normal incidence are plotted in Fig.5a as blue and red lines respectively.

Reflection measurements as a function of the angle of incidence for several wavelengths in the spectral region covering both electric and magnetic dipole resonances are presented as solid circles in Fig.5b for *p*- (red) and *s*-polarized light (blue), together with results of numerical simulations (corresponding solid lines). The best agreement with the experiment was achieved for simulated diameter $D = 170$ nm, height $H = 160$ nm, pitch $P = 300$ nm and substrate refractive index of 1.45. The origin of the small discrepancy between the experiment and simulations is due to a difference between the refractive index of the fabricated silicon and the tabulated data for $\alpha$-silicon[37] used in the simulations, suggesting that the fabricated silicon has less dissipation than that commonly found in literature (see Section 6 in Supplementary Information for details). For *p*-polarization, it is clearly observed the appearance of a zero reflection angle showing strong wavelength dependence and ranging from about 25 to nearly 70 degrees in the studied frequency range, i.e. going well below 45 degrees. For those values below 45 degrees no sign of total internal reflection is found. These results are strongly different from conventional Brewster's angle behaviour and represent the first experimental demonstration of the generalized Brewster effect in arrays of particles with both electric and magnetic responses, i.e., of Brewster-Kerker effect. Numerical simulations, shown as solid lines in Fig.5c, closely reproduce the experimental values and demonstrate excellent agreement in the position of the minima. The slight differences in the reflection intensity, as mentioned, are due to the smaller absorption of the deposited silicon compared to the common amorphous silicon data from literature used in simulations[37]. Taking in simulations slightly lower value of imaginary part of refractive index than in Palik[37] leads to almost perfect agreement to the experiment (see Section 6 in Supplementary Information).

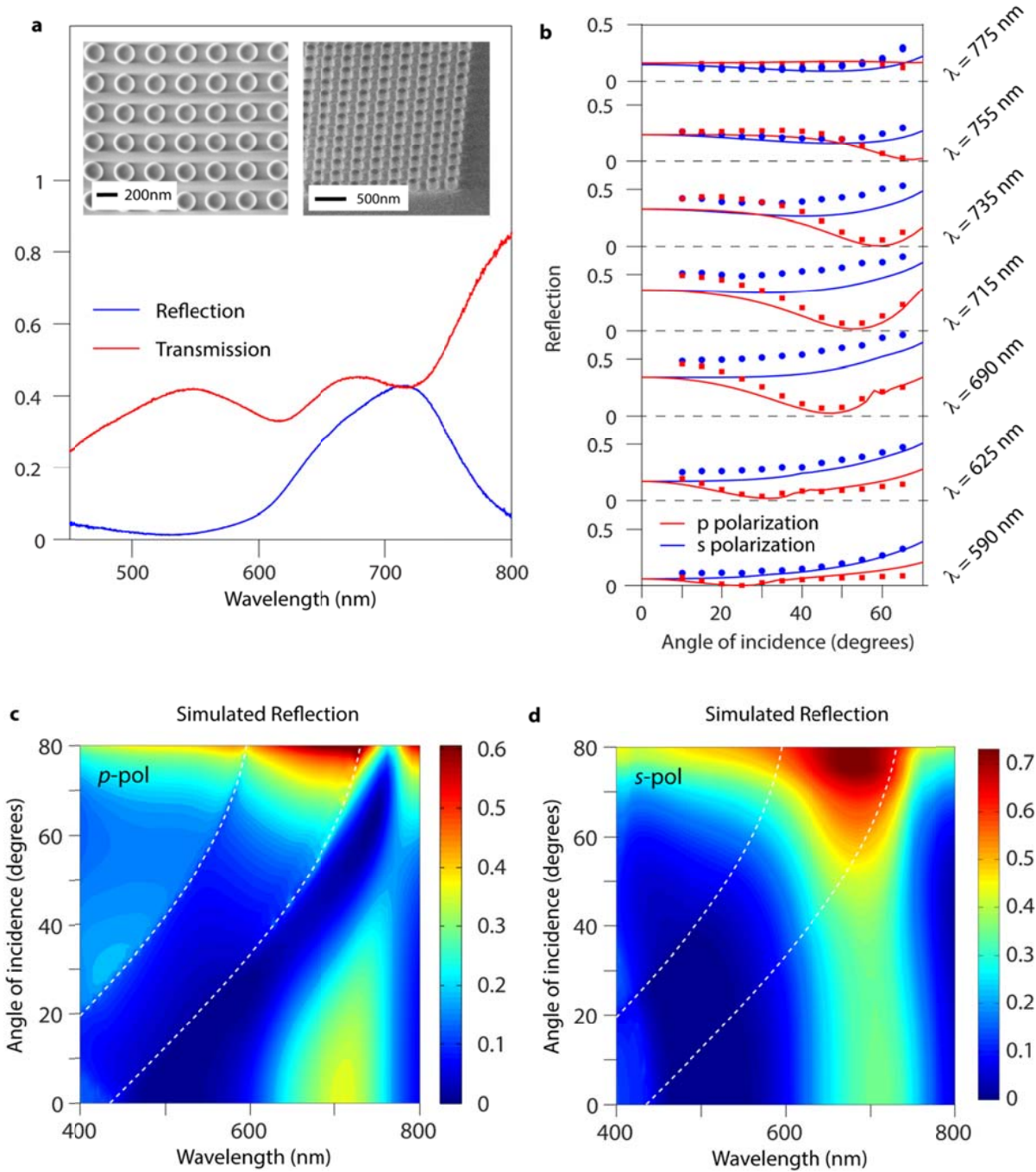

**Figure 5. Experiment and simulations of angular reflection of light from a square lattice of silicon nanodisks over a glass substrate. a,** Experimentally measured reflection (blue solid curve) and transmission (red solid curve) of a square lattice of silicon disks with diameter $D = 180$ nm, height $H = 150$ nm and pitch $P = 300$ nm under normal incidence. The insets show the top (left) and tilted (right) SEM images of the measured sample. **b,** Reflection versus angle of incidence measured for different wavelengths under *p*-polarized (red circles) and *s*-polarized (blue circles) illumination. The corresponding simulated data, obtained for $D = 170$ nm and $H = 160$ nm with the same pitch are shown as solid curves. **c-d,** Maps of simulated values of reflection as a function of angle of incidence and wavelength for *p*-polarized and *s*-polarized incidence, respectively. White dashed lines indicate the onset of the different diffraction orders.

To complete the picture, the full simulated maps of reflection of *p*- and *s*-polarized light as a function of angle of incidence and wavelength are shown in Figs.5c and d. Although with some differences, the general trend observed in the simulated region strongly resembles that shown in Figs.3b and 4a for spheres. For *p*-polarized light, the minimum in reflection strongly varies both with wavelength and angle of incidence, starting in the blue side of the resonances and moving into the region between them for increasing angles. As in the case of spheres both ED and MD modes retrieved from multipole decomposition for single disk in the array are strongly excited in the regions of zero reflection (see Section 7 in the Supplementary Information). Radiation patterns of these interfering dipoles computed for two of the zero reflection cases in *p*-polarization are shown in Fig.S10 (Section 8 in Supplementary Information). They demonstrate vanishing intensities of the radiation in the direction of the reflected wave at the operation wavelength, thus confirming the interference origin of the observed effect.

For *s*-polarization a shallow minimum in reflection at 735 nm, 755 nm and 775 nm can be observed both in simulations and experiment (Fig.5b). These minima correspond to the tail of the vanishing reflection region (see Fig.5d) and provide further experimental evidence of the generalized Brewster effect. An experimental plot focused on the cases of 755 nm and 775 nm can be found in Fig.S9c (Section 7 in the Supplementary Information), clearly showing a minimum in reflection for angles below and above 45 degrees. It is worth mentioning that for this particular system the complete vanishing of reflection under *s*-polarized incidence can be obtained in the spectral region around 850 nm. However, at these wavelengths the array has very low reflectivity even at normal incidence.

Remarkably, even for the realistic system described above the Brewster-Kerker effect is very robust and can easily be detected in experiment, the only true requirement being the efficient excitation of electric and magnetic dipole resonances in the particles forming a sub-diffractive array.

### 3. Conclusions

It has been shown that sub-diffractive arrays of high permittivity dielectric nanoparticles supporting both electric and magnetic dipole resonances present a form of generalized Brewster effect, named Brewster-Kerker effect, leading to vanishing reflection at particular wavelengths and angles both under *p*- and *s*-polarized incidence. The phenomenon can be explained in terms of radiation interference between the electric and magnetic dipoles induced in each particle in the array and connects the angle-suppressed scattering from magneto-electric particles (usually studied in relation to first Kerker's condition) with the zero reflection (Brewster effect) observed in two-dimensional arrangements of such particles. As a consequence of this interference the range of zero reflection angles spans almost over the entire 0-90 degrees without implying total internal reflection. It shows a strong dependence on the incident wavelength and is present for both *p* and *s* polarizations. The effect has been experimentally demonstrated in dense arrays of silicon disks over a fused silica substrate, with measured zero reflection angles ranging from 20 to 70 degrees for wavelengths varying from 590 nm to 775 nm in the visible spectrum. These results represent the first experimental demonstration of the generalized Brewster's effect at optical frequencies in particle arrays with both electric and magnetic response to incident light.

Since this effect is a universal phenomenon related to the directional interference of resonances excited in the particles, it is foreseen that it will be observed in a variety of systems, provided they present electric and magnetic responses. Moreover, tuning the shape and material properties of the particles may lead to almost-on-demand Brewster's effect with regard to polarization, wavelength and angle of incidence. Taking advantage of the strongly resonant character of the structures may bring opportunities for design of efficient sub-wavelength-thick polarizers with a great degree of freedom.

### Methods.

*Numerical simulations of arrays of silicon spheres in air*

Finite Element Method was used to compute the reflection, transmission and absorption of light from infinite square arrays of Silicon spheres (commercial COMSOL Multiphysics software was used). Experimentally measured values of the refractive index of crystalline silicon, taken from Ref. 37, were used in the simulations. The simulation domain consisted of a single unit cell with Bloch boundary conditions applied in the periodicity directions (*x*- and *y*-axes) to simulate an infinite lattice. The so called scattered field formulation of the problem was used. The exciting field was defined as a plane wave with the electric field in the incidence plane for *p*-polarized light and perpendicular to it for *s*-polarized light. Perfectly Matched Layers were applied in the top and bottom directions to absorb all scattered fields from the system. Additionally two planes, $\Sigma_{\pm}$, perpendicular to the *z*-axis at $z = \pm 450$ nm were used as monitors to compute the reflected and transmitted power. Reflection was computed as the flux of the Poynting vector of the scattered fields in the $\Sigma_-$ plane normalized to the power of the plane wave in the same area. Total fields instead of scattered ones were considered in $\Sigma_+$ to compute transmission. Absorption was computed as the volume integration of the Ohmic losses inside the sphere and normalized in the same way. Conservation of energy leads to $R + T + A = 1$, condition that allows internal check of consistency. These results were also checked by performing the same calculation in CST Microwave Studio, showing excellent agreement.

*Numerical simulations of arrays of silicon disks on substrate*

Simulations of silicon disk arrays over substrate (with interface in $z = 0$ and refractive index $n = 1.45$) were carried out using the same approach as described for spheres. The main difference is that, in this case, Fresnel equations were used to explicitly write the excitation fields in the upper ($z > 0$) and lower ($z < 0$) half spaces. While transmission and absorption are computed in exactly the same way, for reflection calculations one needs to consider the Poynting vector of the scattered fields plus the reflected fields from the substrate. In these simulations the refractive index of amorphous silicon[37] was used to approximate the deposited amorphous silicon in the experiment.

*Multipole decomposition*

Multipole decomposition technique was employed to analyse the different modes being excited in the particles. For particles in an array embedded in air, multipoles can be computed through the polarization currents induced within them:

$$\boldsymbol{J} = -i\omega\varepsilon_0(\varepsilon - 1)\boldsymbol{E},$$

where $\varepsilon$ is the permittivity of the particle and $\boldsymbol{E} = \boldsymbol{E}(\boldsymbol{r})$ is the electric field inside it.

This approach fully takes into account mutual interactions in the lattice[38] as well as the possible presence of a substrate. In particular, a Cartesian basis with origin in the centre of the particles was used in the present work. An accurate description of the radiative properties in this basis involves the introduction of the family of toroidal moments[39-41] and the mean-square radii corrections. The explicit expressions of the multipoles as well as the associated partial scattering cross section can be found in Section 9 of the Supplementary Information.

*Nanodisk array fabrication*

Thin films of amorphous silicon of desired thickness were deposited on fused silica substrates via electron beam evaporation (Angstrom Engineering Evovac). The samples were then patterned by single-step electron beam lithography: by spin-coating HSQ resist (Dow Corning, XR-1541-006) and a charge-dissipation layer (Espacer 300AX01), e-beam patterning of the resist (Elionix ELS-7000), and subsequent etching via reactive-ion-etching in inductively coupled plasma system (Oxford Plasmalab 100). The remaining HSQ resist (~ 50 nm after etching) on the top of the nanodisks was not removed since its optical properties after e-beam exposure are close to that of silicon dioxide. To reduce losses the fabricated sample was annealed in vacuum at 600ºC for 40 minutes by using Rapid Thermal Process system (Model: JetFirst200).

*Optical Measurements*

Transmission and reflection measurements of the nanodisk arrays at normal incidence were conducted using an inverted microscopy setup (Nikon Ti-U). For transmission measurements, light from a broadband halogen lamp was normally incident onto the sample from the substrate side before being collected by a 5x objective (Nikon, NA 0.15) and routed to a spectrometer (Andor SR-303i) with a 400 x 1600 pixel EMCCD detector (Andor Newton), as described in detail elsewhere[15]. Transmitted light through the array was normalized to the transmitted power through the substrate only, after accounting for photodetector noise effects (dark current subtraction). For reflection measurements, light from the broadband halogen lamp was incident into the nanodisk array directly passing through the 5x objective. The reflected light was then collected by the same objective and routed into the spectrometer. Reflected light from the array was normalized to the incident power, which is characterized by the reflection of a silver mirror with known spectral response.

Angular transmission and reflection measurements were performed using a home-built free-space microscopy setup. Light originating from a supercontinuum source (SuperK Power, NKT Photonics) was transmitted through a variable band-pass filter for wavelength selection (SuperK Varia, NKT Photonics) and then through a broadband polarizing beam-splitter cube (Thorlabs, PBS252). The linear polarized light passed then through a quarter wave plate (Thorlabs, WPQ10M-808) to obtain circularly polarized light, which was sent to a rotating linear polarizer (Thorlabs, LPNIRE100-B) to obtain linearly polarized light of selected direction. A biconvex lens with 75 mm focusing distance (Thorlabs) was used to focus the light onto the sample surface with silicon nanoparticle arrays. The sample was mounted on a rotation stage for adjusting the angle of incidence. The beam spot size at the sample at normal incidence had a diameter of around 50 μm being smaller than the size of the fabricated arrays (100 μm × 100 μm). A white light lamp source was also coupled into the beam path through the same broadband polarizing beam-splitter cube for sample imaging. Both the incident beam power and the transmitted/reflected beam power were measured by a pixel- photodetector attached to a digital handheld laser power/energy meter console (Thorlabs, PM100D). A scheme of the experimental setup is included in Figure S11 in section 10 of the Supplementary Information.

**Acknowledgements**

Authors from DSI were supported by DSI core funds and A*STAR SERC Pharos program (Singapore). Fabrication and Scanning Electron Microscope imaging works were carried out at the SnFPC cleanroom facility at DSI (SERC Grant 092 160 0139). The authors are grateful to Yi Yang (DSI) and Seng Kai Wong for help with SEM imaging. The work of AEM was supported by the Australian Research Council via Future Fellowship program (FT110100037).

**Author Contributions**

RPD proposed the explanation of the experimental results, performed the theoretical analysis and numerical simulations. YFY performed the sample nanofabrication and the reflection and transmission angular measurements. AIK and LAK built the angular optical measurement setup and performed early stage measurements. AEM performed early stage simulations and developed the phased array model. FYH helped to perform angular and normal-incidence measurements. VV performed SEM imaging. LG, YTT, and AYSK contributed to development of various nanofabrication procedures. BL performed the analysis of the homogeneous slab. AIK proposed the initial idea, and supervised the work. All authors contributed to the manuscript preparation and reviewed the final version of the manuscript.

The authors declare no competing financial interests.

# Supplementary Information

# Generalized Brewster-Kerker effect in dielectric metasurfaces

*Ramón Paniagua-Domínguez[1]\*†, Ye Feng Yu[1]†, Andrey E. Miroschnichenko[2], Leonid A. Krivitsky[1], Yuan Hsing Fu[1], Vytautas Valuckas[1,3], Leonard Gonzaga[1], Yeow Teck Toh[1], Anthony Yew Seng Kay[1], Boris Luk'yanchuk[1], and Arseniy I. Kuznetsov[1]‡*

[1]Data Storage Institute, A*STAR (Agency for Science, Technology and Research),
5 Engineering Drive 1, 117608, Singapore

(corresponding authors: *ramon-paniag@dsi.a-star.edu.sg; ‡arseniy_k@dsi.a-star.edu.sg)

[2]Nonlinear Physics Centre, Research School of Science and Engineering, The Australian National University, Acton, ACT, 2601, Australia

[3] Department of Electrical and Computer Engineering, National University of Singapore,
1    Engineering Drive 2, 117576, Singapore

## 1. Generalized Brewster's effect for a magneto-electric slab.

The present section aims to present the not-so-well-known rich phenomenology associated with reflection of plane waves at an interface between an ordinary medium and one having simultaneous electric and magnetic responses. Instead of analyzing a single interface, let us focus on the case of a slab, located either in air or standing over a semi-infinite glass, since these cases arguably model more accurately the system studied in the main manuscript.

Consider the general case represented in Fig.S1. The thickness of the film with optical properties given by $\varepsilon_2$ and $\mu_2$ is $h$, while media 1 and media 3 are semi-infinite. As usual, two different polarizations (*p*- and *s*-) should be considered.

Similar to the case of ordinary dielectrics, Fresnel's formulas follow from the boundary conditions of electric and magnetic fields at interfaces $z=0$ and $z=h$ (continuity of the tangential field components). This yields four equations for the *y*-components of the magnetic or electric fields (for *p* or *s*-polarization, respectively). Solving these equations the formulas for the amplitudes of the reflection coefficients are obtained.

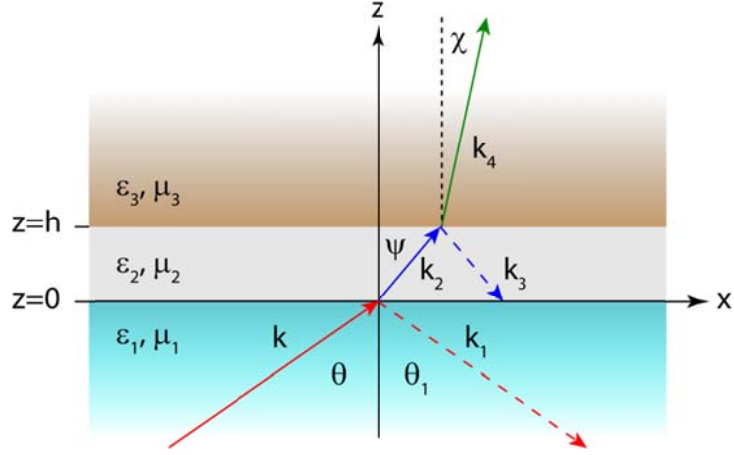

**Figure S1.** Schematics of light transmission and reflection from a slab confined between two infinite media.

a) ***p-polarization***. The reflection coefficient is given by $r_p = H_{1y}/H_{0y}$, where $H_{0y}$ and $H_{1y}$ are the magnetic field amplitude of the forward (incident) and backward (reflected) propagating waves in media 1. The reflectivity is given by $R_p = |r_p|^2$. This formula can be presented in standard form[1]:

$$r_p = \frac{r_{12}^{(p)} e^{-2i\psi} + r_{23}^{(p)}}{e^{-2i\psi} + r_{12}^{(p)} r_{23}^{(p)}}, \qquad (1S)$$

where the amplitude reflection coefficient $r_{12}$ (between media 1 and media 2) is given by:

$$r_{12}^{(p)} = \frac{\varepsilon_2 \sqrt{\mu_1} \cos\theta - \sqrt{\varepsilon_1}\sqrt{\varepsilon_2\mu_2 - \varepsilon_1\mu_1 \sin^2\theta}}{\varepsilon_2 \sqrt{\mu_1} \cos\theta + \sqrt{\varepsilon_1}\sqrt{\varepsilon_2\mu_2 - \varepsilon_1\mu_1 \sin^2\theta}} \qquad (2S)$$

and the amplitude reflection coefficient $r_{23}$ (between media 2 and media 3) is given by:

$$r_{23}^{(p)} = \frac{\varepsilon_3 \sqrt{\varepsilon_2\mu_2 - \varepsilon_1\mu_1 \sin^2\theta} - \varepsilon_2 \sqrt{\varepsilon_3\mu_3 - \varepsilon_1\mu_1 \sin^2\theta}}{\varepsilon_3 \sqrt{\varepsilon_2\mu_2 - \varepsilon_1\mu_1 \sin^2\theta} + \varepsilon_2 \sqrt{\varepsilon_3\mu_3 - \varepsilon_1\mu_1 \sin^2\theta}} \qquad (3S)$$

and $\psi = \psi(\theta)$ represents the change in the wave phase over the thickness $h$ of the layer ($k_0 = \omega/c$):

$$\psi = k_0 h \sqrt{\varepsilon_2\mu_2 - \varepsilon_1\mu_1 \sin^2\theta} \qquad (4S)$$

**b)** *s-polarization*. In a totally analogous way, the reflection coefficient is given by $r_s = E_{1y}/E_{0y}$, and the reflectivity is $R_s = |r_s|^2$. In this case:

$$r_s = -\frac{r_{12}^{(s)} e^{-2i\psi} + r_{23}^{(s)}}{e^{-2i\psi} + r_{12}^{(s)} r_{23}^{(s)}}, \tag{5S}$$

where:

$$r_{12}^{(s)} = \frac{\mu_2 \sqrt{\varepsilon_1} \cos\theta - \sqrt{\mu_1} \sqrt{\varepsilon_2 \mu_2 - \varepsilon_1 \mu_1 \sin^2\theta}}{\mu_2 \sqrt{\varepsilon_1} \cos\theta + \sqrt{\mu_1} \sqrt{\varepsilon_2 \mu_2 - \varepsilon_1 \mu_1 \sin^2\theta}}, \tag{6S}$$

$$r_{23}^{(s)} = \frac{\mu_3 \sqrt{\varepsilon_2 \mu_2 - \varepsilon_1 \mu_1 \sin^2\theta} - \mu_2 \sqrt{\varepsilon_3 \mu_3 - \varepsilon_1 \mu_1 \sin^2\theta}}{\mu_3 \sqrt{\varepsilon_2 \mu_2 - \varepsilon_1 \mu_1 \sin^2\theta} + \mu_2 \sqrt{\varepsilon_3 \mu_3 - \varepsilon_1 \mu_1 \sin^2\theta}}, \tag{7S}$$

Note that amplitude coefficients of reflection $r_p$ and $r_s$ in the limiting case $\theta = 0$ differ in sign, as **E** represents a polar, and **H** an axial vector[2].

From this analysis it is readily seen that with appropriate variations of $\varepsilon$ and $\mu$ it is possible to obtain arbitrary values for the Brewster angle, corresponding to the vanishing value of reflectivity, both for *p*-polarized light and for *s*-polarized light.

We illustrate now the phenomenology associated with the generalized Brewster's effect by considering a slab with $\varepsilon$ and $\mu$ standing in air. Results, shown in Fig.S2, are selected to illustrate the main characteristics of the generalized Brewster, namely, the possibility to obtain Brewster angle for s-polarized light (Fig.S1a-c) and the possibility to obtain Brewster for angles below 45 degrees without having total internal reflection (TIR) for larger angles (Fid.S2c and f). Note that this phenomenology is analogous to that observed in the case of sub-diffractive silicon nanosphere array embedded in air, as presented in section 1 of the main text.

For the sake of completeness we illustrate, in Fig.S3 the case of a magneto-electric slab on top of a glass semi-infinite medium (incidence from the side of air). As readily seen, this configuration retains all major characteristics, and serves to illustrate the phenomenology found in the case of silicon nanoparticle array over glass substrate presented in section 2 of the main text.

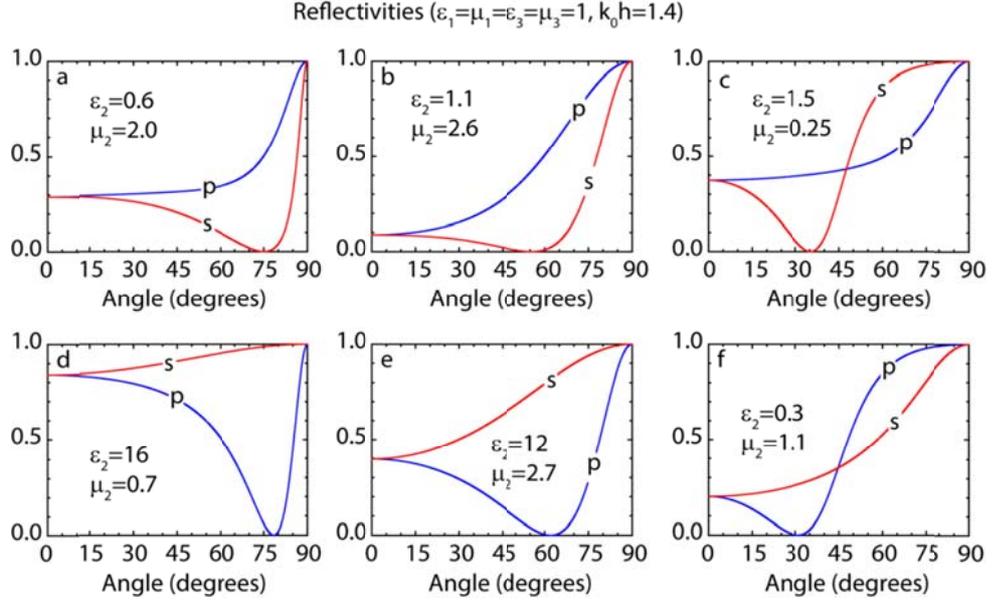

**Figure S2. Reflectivity of a magneto-electric slab in vacuum:** $\varepsilon_1 = \mu_1 = \varepsilon_3 = \mu_3 = 1$ and $k_0 h = 1.4$ (corresponding to $h = 150$ nm at wavelength $\lambda \sim 670$ nm) versus angle of incidence for different values of $\varepsilon_2$ and $\mu_2$. **a-c** Brewster's angle for *s*-polarization; **d-f** Brewster's angle for *p*-polarization. Arbitrary values of Brewster's angle can be obtained between 0° and 90° without having total internal reflection at larger angles.

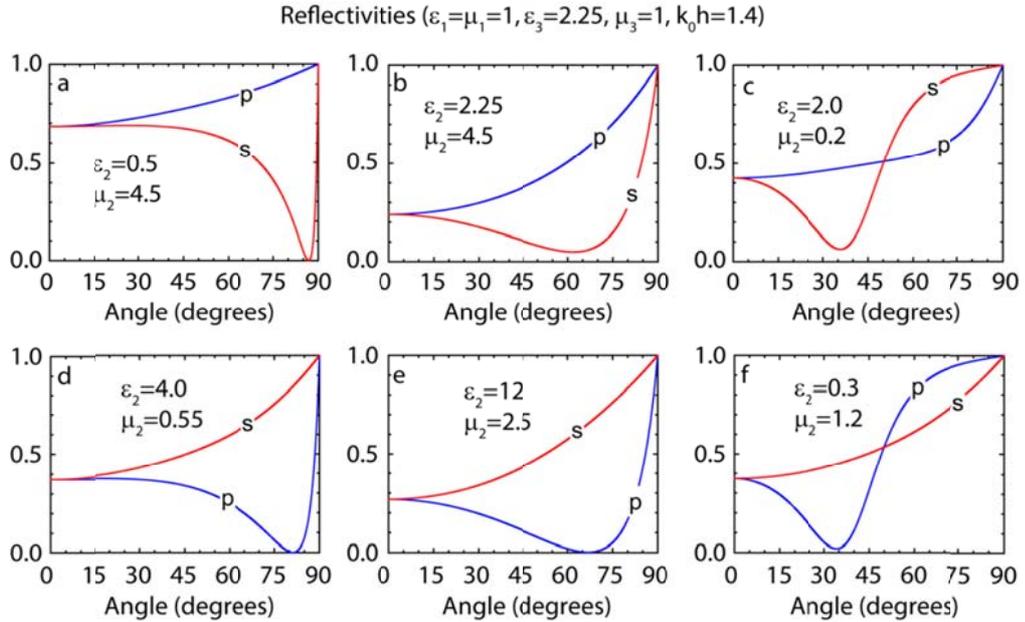

**Figure S3. Reflectivity of a magneto-electric slab on top of glass:** $\varepsilon_1 = \mu_1 = 1$, $\varepsilon_3 = 2.25$, $\mu_3 = 1$ and $k_0 h = 1.4$ (corresponding to $h = 150$ nm at wavelength $\lambda \sim 670$ nm) versus angle of incidence for different values of $\varepsilon_2$ and $\mu_2$. **a-c** Brewster's angle for *s*-polarization; **d-f** Brewster's angle for *p*-polarization. Arbitrary values of Brewster's angle can be obtained (0° - 90°) without having total internal reflection.

## 2. Electric far-field radiated by a pair of electric and magnetic dipoles

Consider a pair of electric and magnetic dipoles. The electric far-field radiated in the direction given by the unit vector $\hat{n}$ can be written as:

$$\boldsymbol{E}_{ff} = \boldsymbol{E}_{ff}{}^p + \boldsymbol{E}_{ff}{}^m = \frac{k_0^2}{4\pi\epsilon_0}\left[\hat{n} \times (\boldsymbol{p} \times \hat{n}) + \frac{1}{c}\boldsymbol{m} \times \hat{n}\right] \tag{8S}$$

with $k_0 = 2\pi/\lambda$ the wavenumber and $\epsilon_0$ and $c$ the permittivity and speed of light in vacuum, respectively.

Consider the situations depicted in Fig.2a of the main text. The induced dipoles, oscillating parallel to the driving incident fields can be written as $\boldsymbol{p} = (-p\cos\theta_i,\ 0,\ p\sin\theta_i)$ and $\boldsymbol{m} = (0,\ m\cdot c,\ 0)$, with $p$ and $m$ the complex amplitudes of the induced dipoles and $\theta_i$ the angle of incidence. In this situation, the radiated (scattered) field in a direction of observation given by the polar angle $\theta$ in the xz-plane (highlighted in the figure) is purely polar and reads:

$$\boldsymbol{E}_{ff} \propto [m - p\cos(\theta - \theta_i)]\hat{\boldsymbol{\theta}}, \tag{9S}$$

with $\hat{\boldsymbol{\theta}}$ the unitary polar vector. It is clear that, in this situation, the electric field is suppressed if:

$$\cos(\theta - \theta_i) = m/p = (|m|/|p|)e^{i\delta} \tag{10S}$$

with $\boldsymbol{\delta}$ being the phase difference between the two dipoles. Whenever the phase difference between dipoles is a multiple of $\boldsymbol{\pi}$ the field exactly vanishes. It is clear that, when an infinite array of spheres is considered in the xy-plane, this situation represents the *p*-polarization incidence case, and the plane of incidence coincides with the xz-plane.

Analogously, when the case depicted in Fig.2b holds, the induced electric and magnetic dipoles can be described by $\boldsymbol{m} = (-m\cdot c\cos\theta_i,\ 0,\ m\cdot c\sin\theta_i)$ and $\boldsymbol{p} = (0, -p,\ 0)$. In this situation, the radiated (scattered) field in the plane containing the magnetic dipole (highlighted in Fig.2b) is purely azimuthal and reads:

$$\boldsymbol{E}_{ff} \propto [p - m\cos(\theta - \theta_i)]\hat{\boldsymbol{\phi}} \tag{11S}$$

with $\hat{\boldsymbol{\phi}}$ the unitary azimuthal vector. It vanishes if:

$$\cos(\theta - \theta_i) = p/m = (|p|/|m|)e^{-i\delta}, \tag{12S}$$

and will represent the *s*-polarization case for infinite arrays.

### 3. Phased arrays of point scatters.

It is known from the phased array antennas theory that the total intensity from an array of identical emitters can be expressed as:

$$I(\theta, \phi) = |F(\psi)|^2 |E_{single}(\theta, \phi)|^2 \tag{13S}$$

where $F(\psi)$ is the so called form factor of the array, which describes the phase retardation from different elements in the lattice and $E_{single}(\theta, \phi)$ is the far-field of each identical constituent. An analogous formula holds to describe the scattering properties of an array of identical point-like scatters. As in the case of phased array antennas, $F(\psi)$ carries information about the geometry of the array and does not depend on the particular scatters considered. It reads:

$$F(\psi) = \frac{\sin\left(\frac{N\psi}{2}\right)}{N \sin\left(\frac{\psi}{2}\right)}$$

in which $\psi = kd \sin\theta \cos\phi + \xi$, with $k = 2\pi/\lambda$ being the wavenumber, $d$ the lattice period and $\xi = kd \sin\theta_i$ the phase difference due to oblique incidence at an angle $\theta_i$ (we consider the plane of incidence as $\phi = 0$). Here $N$ is the number of particles in the array. In the limit $N \to \infty$ one has:

$$\lim_{N \to \infty} F(\psi) = \begin{cases} 1, & \psi \to 0 \\ 0, & \psi \nrightarrow 0 \end{cases}$$

Fixing the scattering plane to $\phi = 0$, $F(\psi)$ is non-zero only when:

$$\sin\theta + \sin\theta_i = 2\sin\left(\frac{\theta+\theta_i}{2}\right)\cos\left(\frac{\theta-\theta_i}{2}\right) = 0.$$

This implies that, if no higher diffracted order are present, $F(\psi)$ is non-zero when $\theta = -\theta_i \equiv \theta_r$ or $\theta = \pi + \theta_i \equiv \theta_t$. In Fig.S4a, $|F(\psi)|^2$ is plotted for $\theta_i = \pi/6$, $\lambda = 730$ nm and $d = 300$ nm for several increasing number of particles $N$. As seen, it quickly converges to the limit above, vanishing everywhere except in the reflection and transmission directions.

Let us now assume that each single element in the array is a pair of electric ($\boldsymbol{p}$) and magnetic ($\boldsymbol{m}$) dipoles. The radiated far-field $\boldsymbol{E}_{single}(\theta, \phi)$ will be given by equation (8S), see section 2 above. Consider the two main situations presented there. In the first the electric dipole is contained in the plane of incidence ($\phi = 0$) with $\boldsymbol{p} = (-p\cos\theta_i, \ 0, \ p\sin\theta_i)$, and $\boldsymbol{m} = (0, \ m \cdot c, \ 0)$. In this case, the radiated field in this plane is given by (9S). Clearly, this situation will represent the case of *p*-polarized incidence. In the second case, the magnetic dipole is contained in the incidence plane and reads $\boldsymbol{m} = (-m \cdot c \cos\theta_i, \ 0, \ m \cdot c \sin\theta_i)$ while $\boldsymbol{p} = (0, -p, 0)$. In this situation the radiated field in the plane of incidence is given by (11S) and will represent the case of *s*-polarized incidence.

From (10S) and (12S) one can compute the relative values of $p$ and $m$ for which the field at $\theta = \theta_r = -\theta_i$ will be zero. In this case, no intensity at all will be radiated in the reflection direction (as follows from equation (13S)), leading to perfect transmission, i.e., to Brewster's effect. Let us consider, e.g. the case of $\theta_i = \pi/6$ and *s*-polarization. From equation (12S) it follows that the field vanishes at $\theta = -\theta_i = -\pi/6$ for $p = m/2$. Note that this relation immediately implies $C_{sca}^{ED}/C_{sca}^{MD} = 1/4$ (see equations (4) and (5) of the main text), which precisely corresponds to the case depicted in the bottom panel of Fig.2c in the main text. In Fig.S4b we plot $|F(\psi)|^2$ (left), $|\boldsymbol{E}_{single}(\theta, \phi)|^2$ (center) and $I(\theta, \phi)$ (right) for this case (with $\theta_i = \pi/6$, $\lambda = 730$ nm, $d = 300$ nm and $N = 500$). It is clear from the calculation that, due to the modulation of the form factor $|F(\psi)|^2$, radiation in any other direction rather than those of transmission and reflection is totally inhibited due to interference from different lattice sites, even if the single particles radiate in those directions. It is also immediately seen that the suppression of radiation in the reflection direction from each single element implies the suppression of radiation from the whole array. Finally, to stress the origin of the effect in the inhibition of radiation from single elements, we plot in Fig.S4c the case $p = m/3$. This ratio does not lead to zero radiation in the reflection direction and, thus, no Brewster is obtained.

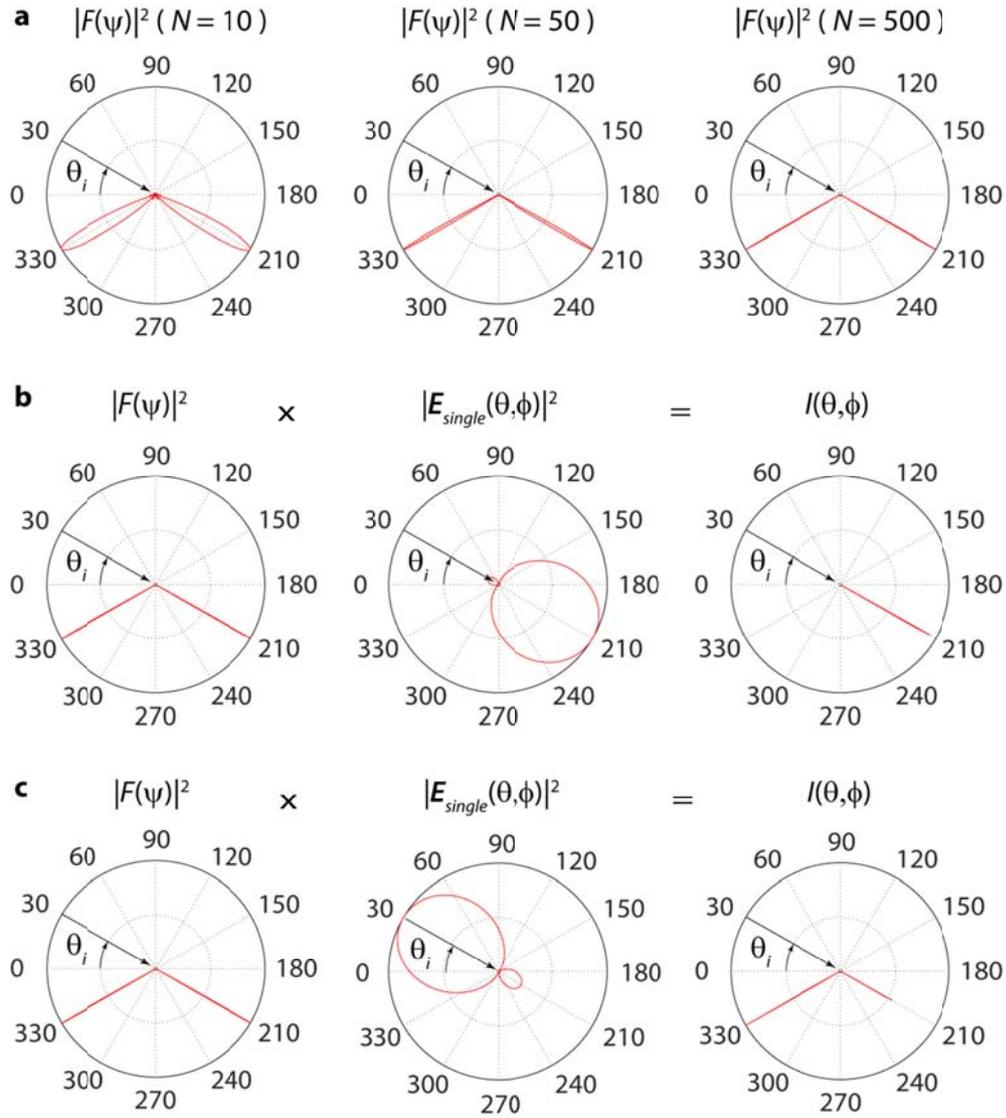

**Figure S4. Scattering properties of a phased array of identical electric and magnetic dipoles. a,** The form factor $|F(\psi)|^2$ of the array as a function of the number of particles considered. The form factor quickly converges towards zero everywhere except at the reflection and transmission direction, for which $|F(\psi)|^2 = 1$. The direction of incidence is indicated by a black arrow together with the incidence angle $\theta_i = \pi/6$. The wavelength is $\lambda = 730$ nm and the pitch $d = 300$ nm. **b,** The form factor $|F(\psi)|^2$ (left), the radiation pattern of each single element in the array $\boldsymbol{E}_{single}(\theta, \phi)$ (center) and the total radiated intensity from the array $I(\theta, \phi)$ (right) for an array of electric and magnetic dipoles with $\boldsymbol{m} = (-m \cdot c \cos\theta_i,\ 0,\ m \cdot c \sin\theta_i)$ and $\boldsymbol{p} = (0, -p, 0)$, fulfilling the condition $p = m/2$. Parameters $\lambda$, $\theta_i$ and $d$ are the same as in **a**, $N = 500$. This configuration suppresses reflection in *s*-polarization. **c,** The same as in b but with $p = m/3$ for which reflection is not suppressed.

## 4. Absorption and higher order multipoles in arrays of spheres at *p*-polarized incidence.

It is our intention here to complete the picture given in Section 1 of the main manuscript, regarding the analysis of the resonances excited in the array of silicon (Si) spheres with diameter D = 180 nm and pitch P = 300 nm for different wavelengths and angles of incidence for *p*-polarized light. As mentioned in the main text, the electric and magnetic dipole contributions are the dominant ones in the range of wavelengths and angles of incidence studied. Those are shown in the whole simulated range in Fig.S5a and b, respectively. Also, the electric quadrupole partial scattering cross section, as computed through the multipole decomposition, is shown in Fig.S5b, while the corresponding plot for the magnetic quadrupole is shown in Fig.S5c. As readily seen, both resonances appear for wavelengths much shorter than those for which the generalized Brewster effect is observed. Figure S5d also shows the absorption in the array, computed through volume integration of the Ohmic losses inside the spheres. These results serve first as a demonstration of the energy conservation in our simulations and also to track all resonances excited in the system, showing excellent correspondence with those computed through the multipole decomposition technique (the magnetic one, however, is fainter due to the lower dissipation of silicon at that wavelength).

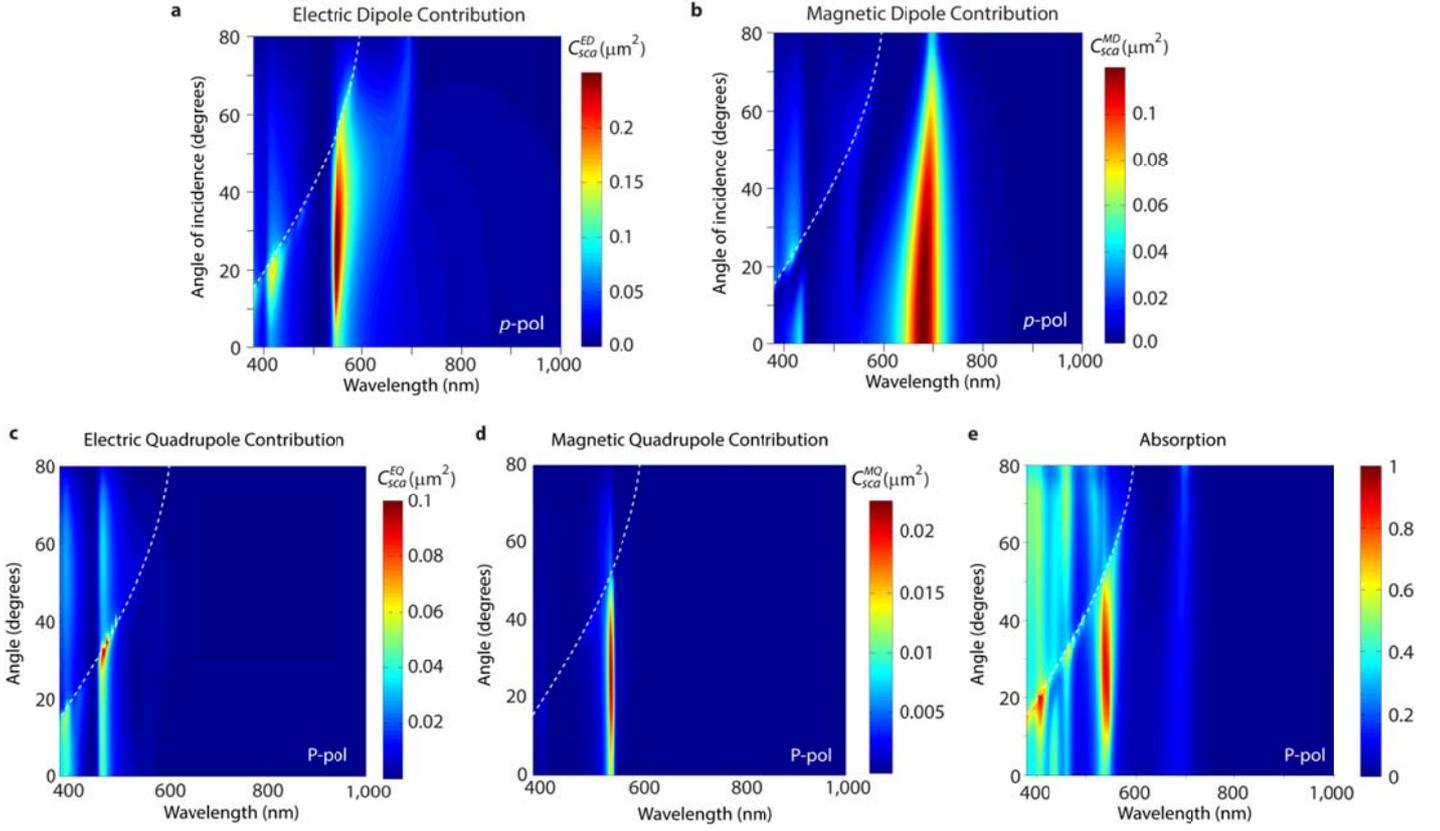

**Figure S5. Multipolar contributions and absorption in a square lattice of silicon spheres versus wavelength and angle of incidence under *p*-polarized incidence. a,** and **b,** respectively, electric and magnetic dipolar contributions to the scattering from each particle in the array as retrieved by multipole decomposition. Dashed white lines show the spectral position of the first diffraction order. **c** and **d**, respectively, electric and magnetic quadrupolar contributions to the scattering from a particle in the array as retrieved by multipole decomposition. **e**, Absorption computed through volume integration of the Ohmic losses.

5. **Interpretation of the observed generalized Brewster effect based on the scattering characteristics of a single sphere.**

Let us consider more carefully the scattering characteristics of a single silicon sphere, Fig.S6 (the figure is similar to Fig.2b in the main manuscript, but some points of interest are added). Although the real picture is more complicated due to the inter-particle interactions inside the arrays, a simplified model of a single sphere may reveal major features of the reflection maps in Figs.3b and 4a from the main manuscript. In this spirit let us assume here that particles in the lattice behave similar to individual scatters. In this case, as already mentioned, cancellation of radiation in the plane containing the incident electric field (corresponding to the case of *p*-polarized incidence) is only possible when the induced electric dipole dominates, i.e., in the red shaded regions in Fig.S6. Cancellation in the perpendicular plane (corresponding to the case of *s*-polarized incidence) is only possible under the blue shaded regions (dominance of magnetic dipole). Since the scattering angle and the reflection angle are related as $\theta = -\theta_i$, it is clear that for $\theta_i < 45$ the dipoles must be in phase (yellow shaded regions) while for $\theta_i > 45$ they must be in anti-phase (green shaded region).

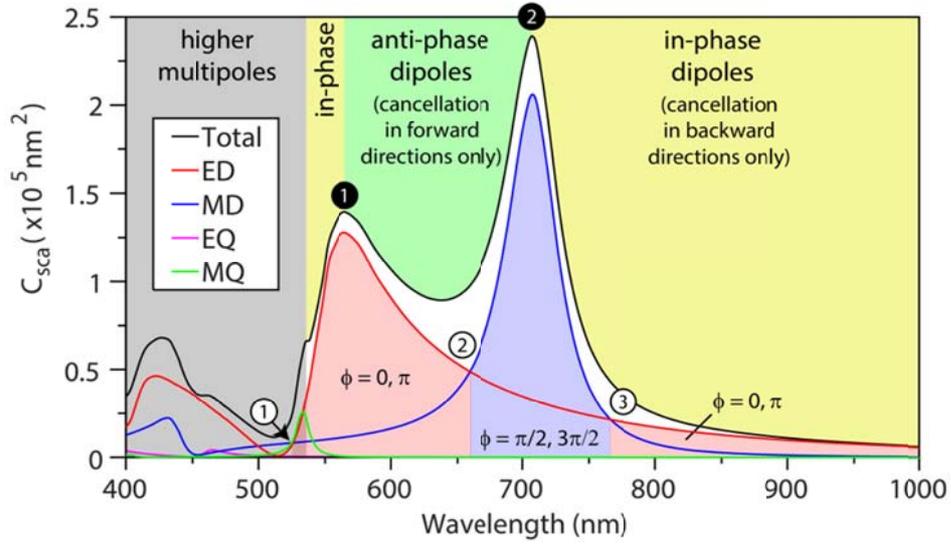

**Figure S6. Optical properties of a silicon sphere with diameter $D = 180$ nm in air under plane wave illumination, and regions and points of interest.** Total scattering cross section (black curve) and contributions from the electric dipole (red curve), magnetic dipole (blue curve), electric quadrupole (magenta curve) and magnetic quadrupole (green curve).

Let us start by analyzing the *p*-polarized case. Thus, to achieve the zero reflection effect we must restrict ourselves to the red shaded regions where electric dipole dominates. At normal incidence the first Kerker's condition, indicated by ① and ③ in Fig.S5, leads to zero reflection. Let us first analyze the spectral region around ①. For increasing angles of incidence, equation (4) implies that $C_{sca}^{MD}/C_{sca}^{ED}$ should correspondingly decrease. This is achieved at longer wavelengths with respect to ①, which manifests as the slight redshift in the zero of reflection in Fig.3b for angles below 45 degrees. Above 45 degrees, the dipoles have to be in opposite phases to cancel radiation in the reflection direction, thus crossing ❶ in Fig.S6 and entering in the green region, as it is observed in the zero of reflection in Fig.3b. In order to satisfy equation (4) now the rate $C_{sca}^{MD}/C_{sca}^{ED}$ should increase instead, which is again possible at longer wavelengths. In this region, however, the range of wavelengths is wider going up to ② (above ② MD contribution starts to dominate), leading to a more pronounced redshift in Fig.3b. Thus, the sequence ① → ❶ → ② always implies a redshift to fulfill equation (4), as observed in Fig.3b. Interestingly, if now ③ is chosen as the starting point, fulfilling equation (4) again implies longer wavelengths for larger angles of incidence. However, cancellation is only possible below 45 degrees, since there is no region in which the dipoles are in anti-phase, thus explaining the asymptotic behavior of the zero in reflection < 45 degrees observed in Fig.3b.

Having analyzed the *p*-polarized case, the corresponding analysis of *s*-polarization is straightforward. We are now restricted to move within the blue shaded region. Starting again in Kerker's first condition at normal incidence ③, fulfilling equation (5) now implies a blue-shift. Since at 45 degrees the dipoles must change from in-phase to anti-phase, the complete sequence is now ③ → ❷ → ②, which implies a constant blue-shift in the whole range, as observed in Figs.4a and b. Since the blue area is narrower, this directly translates in a narrow spectral band for zero reflection, which in the real system gets reinforced by a narrowing of the magnetic resonance due to the lattice interactions.

Let us conclude showing that the main features observed in the reflectivity of the arrays can be obtained in a simple way from equations (8S)-(12S), which describe the radiation of a pair of electric and magnetic dipoles. For that, let us assume that the electric ($\alpha_E$) and magnetic ($\alpha_M$) polarizabilities of the dipoles are those of a Si sphere according to Mie theory (i.e., $\alpha_E = 6\pi i a_1/k^3$ and $\alpha_H = 6\pi i b_1/k^3$, with $a_1$ and $b_1$ the electric and magnetic dipolar scattering coefficients, respectively[3]). One important assumption is made to correctly reproduce the results. The dipoles are assumed to change their phase abruptly around the resonance peak. For single spheres this only holds approximately but it correctly models the effect of interactions in the array. Of course one could fully take into account the effect of the lattice by computing the self-consistent field at each dipole position and computing the effective polarizabilities. However, it is enough to have good results to consider that the effect of the lattice manifests just as a steeper phase change in the polarizabilities of the particles. Thus, we take the amplitudes as given by Mie theory but assume a step function for the phases, as depicted in Fig.S7a. Then, by simply applying equations (9S) and (11S), it is possible to compute the intensity radiated in the reflection angle $\theta = -\theta_i$, as a function of the angle of incidence $\theta_i$ and wavelength $\lambda$ for *p*- and *s*-polarization, respectively. Recall that the condition of zero radiated intensity in the reflection direction is sufficient to have zero reflection (see section 3). The computed intensity is shown in Figs.S7b and c for *p*- and *s*-polarizations. It can be seen that the zones showing zero radiated intensity in the reflection direction closely reproduce those of zero reflectivity found by full numerical simulations (Figs.3b and 4a). For *p*-polarization, it correctly predicts the red-shift of the region of zero reflection at long wavelengths (red side of resonances) as it approaches $\theta_i = 45$ degrees. Also, it reproduces the behavior of the zero reflection region at shorter wavelengths (blue side of the resonances). It predicts its continuous red-shift for increasing angles of incidence and the "jump" to the region between resonances for $\theta_i > 45$. Finally, for s-polarization, it correctly predicts the continuous blue-shift of the zero intensity region located in the red side of the resonances.

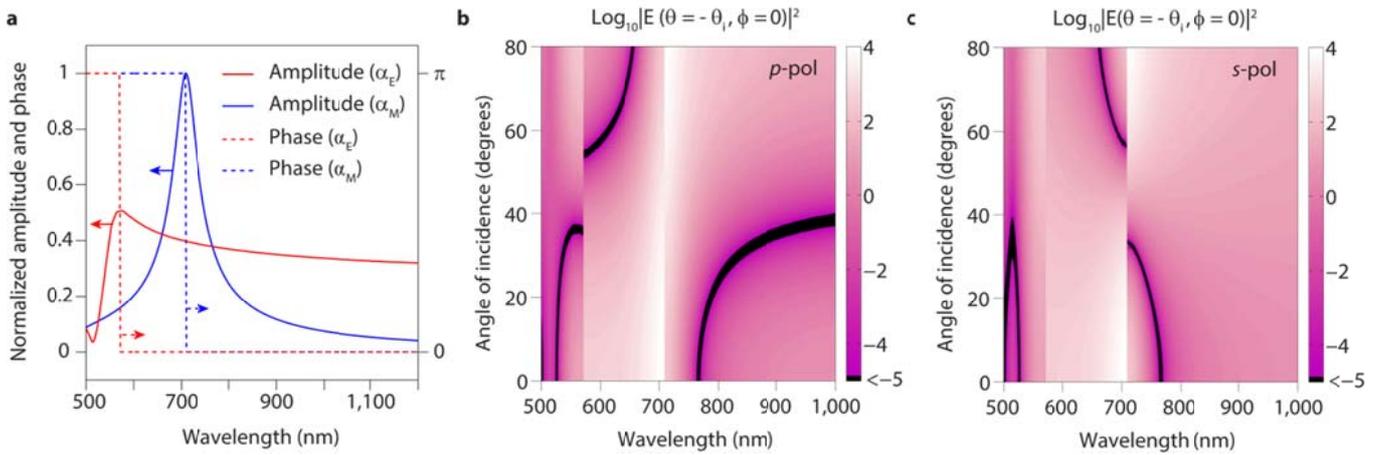

**Figure S7. Radiated intensity in the reflection direction by a pair of electric and magnetic dipoles under *p*-polarized (red) and *s*-polarized (blue) light. a,** Amplitudes and phases of the polarizabilities of the dipoles. **b,** Radiated intensity in the plane of incidence ($\phi = 0$) and reflection direction ($\theta = -\theta_i$) by the pair of dipoles under *p*-polarization. **c,** The same as in **b** but for *s*-polarization.

# 6. Impact of losses on the angular reflection of a square lattice of nanodisks on top of fused silica. Experiment vs simulations.

In the present section we demonstrate that, as mentioned in the main text, the differences observed between experiment and simulations in the angular reflection of a sample of silicon nanodisks on top of silica substrate are almost entirely due to the lower absorption of the fabricated sample compared to values tabulated[4] for amorphous silicon (a-Si).

As seen in Fig.5 in the main text, the differences are more pronounced above 600nm. Below that limit the agreement is fairly good (see the case at 590nm). Above, however, experiment and theory quickly depart, and reflection is higher in the fabricated sample, indicating a quick drop of absorption as compared to the tabulated data used for simulations.

In Fig.S8 we show the same set of curves as in Fig.5d of the main manuscript but, instead of directly taking the complex refractive index ($n' + in''$) from Ref.4 we take only the real part ($n'$), and allow the imaginary ($n''$) to be smaller. For each wavelength, we choose it in such a way that measured values show good agreement for low angles of incidence (thus fitting the spectrum at normal incidence). It is readily seen that the agreement between experiment and simulations obtained in this way is excellent.

In Table S1 we show the set of values of $n''$ used together with those tabulated. At 590 nm we just took the same value as the tabulated. The mentioned quick drop of the absorption is clearly seen and the value at 775 nm approaches that of Ref.4.

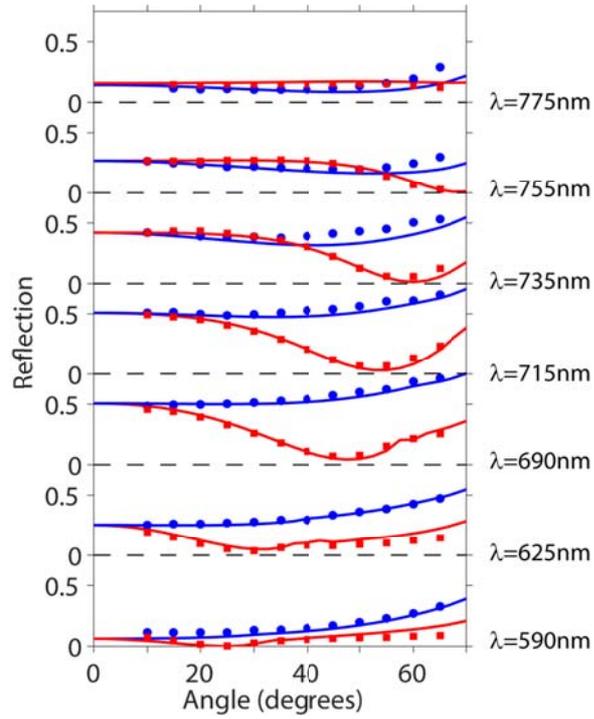

**Figure S8. Reflection vs angle of incidence under *p*-polarized (red) and *s*-polarized (blue) light in a square lattice of a-silicon nanodisks over a fused silica substrate.** The plot shows the comparison between measured values (symbols) and simulations (lines) obtained fitting the imaginary part of the index ($n''$). Real part ($n'$) was taken from Ref.4.

**Table S1.** Values of imaginary part of refractive index used in simulations in Fig.S8 and those from Ref.4.

| Wavelength (nm) | $n''$ (fitting) | $n''$ (Ref.4) |
|---|---|---|
| 590 | 0.583 | 0.583 |
| 625 | 0.300 | 0.445 |
| 690 | 0.140 | 0.269 |
| 715 | 0.135 | 0.224 |
| 735 | 0.130 | 0.191 |
| 735 | 0.130 | 0.164 |
| 755 | 0.130 | 0.136 |

# 7. Multipole contributions and absorption in the square lattice of Si nanodisks on top on fused silica substrate.

In order to complete the analysis of the generalized Brewster's effect for the Si nanodisks metasurface with pitch $P = 300$ nm, diameter $D = 170$ nm and height $H = 160$ nm, given in section 2 of the main manuscript, we present here some additional results. In particular, the electric and magnetic dipolar contributions to the scattering from a single element in the array as a function of wavelength and angle of incidence under irradiation with p-polarized light are shown in Fig.S9a and b. Also, we plot in Fig.S9c the reflection of s-polarized light as a function of angle of incidence for the particular cases of $\lambda = 755$ nm and $\lambda = 775$ nm. It is readily observed the emergence of a minimum in reflection, the angle of which decreases with increasing wavelength. Note that this angle can have values above (for $\lambda = 755$ nm) and below ($\lambda = 775$ nm) 45 degrees, a clear signature of the generalized Brewster's effect. We also include, for completeness, the absorption, both under s- and p-polarized incidence, in Figs.S9d and e, respectively. It allows tracking the resonances excited in the system. For p-polarized light, comparison with Figs.S9a and b also serves as a verification of the resonances computed by the multipole decomposition.

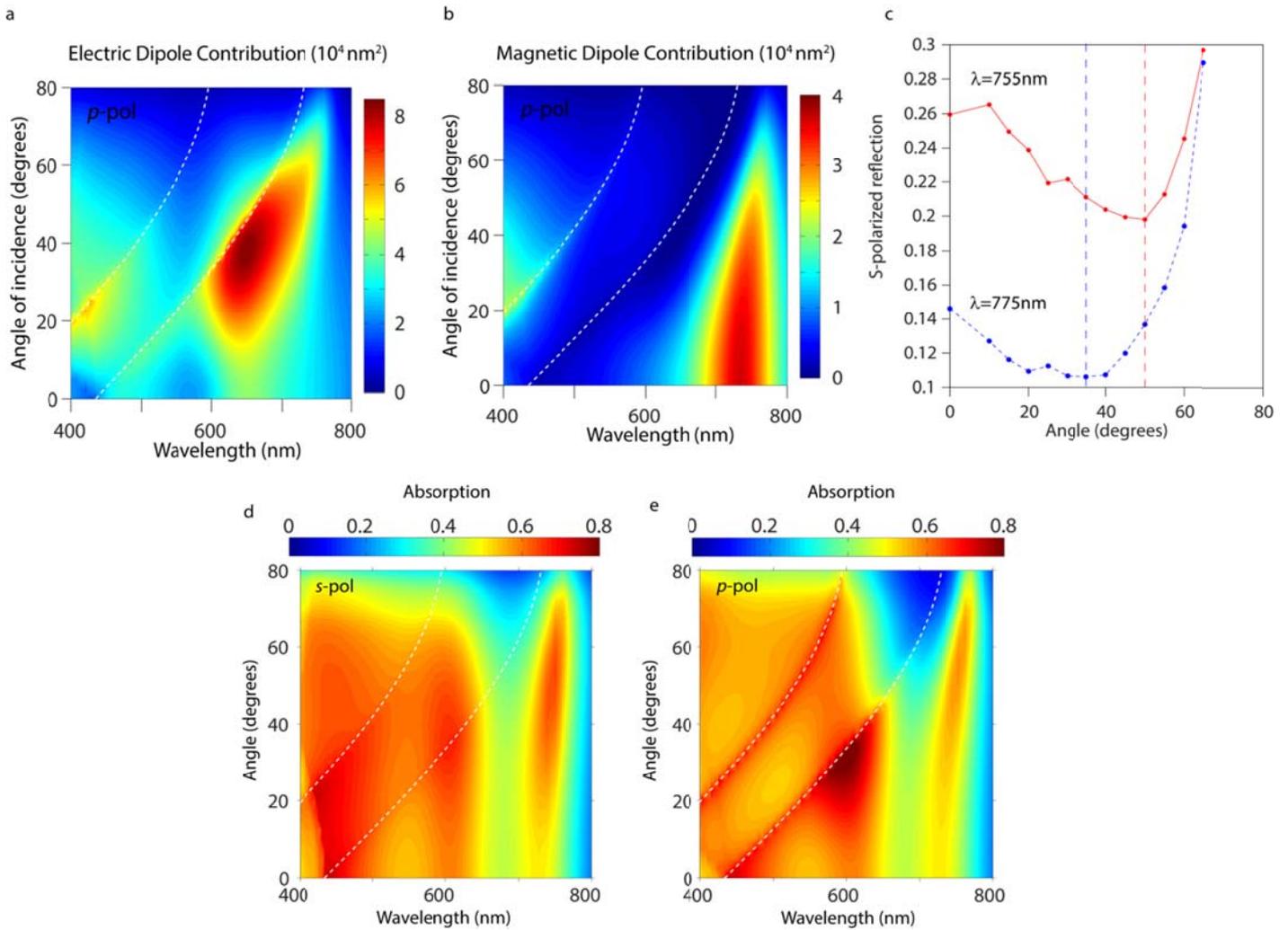

**Figure S9. Multipolar contributions from a single disk in the array under *p*-polarized incidence, detail of reflection under *s*-polarized light and absorption in both polarizations in a square lattice of silicon nanodisks over a fused silica substrate versus angle of incidence and wavelength. a-b,** Simulated electric (a) and magnetic (b) dipole contributions to the total scattering cross section of a single disk in the square lattice of silicon nanodisks over a fused silica substrate under p-polarized light. **c,** Details of the measured angular reflection from the array under s-polarized light for particular wavelengths. **d, e,** Simulated absorption computed through volume integration of the ohmic losses inside the particles, under *s*- and *p*-polarized incident light, respectively.

## 8. Radiation patterns of each nanodisk in a square lattice on top of fused silica substrate leading to vanishing reflection at p-polarized incidence.

In this section, the radiation patterns emitted by pairs of electric and magnetic dipoles excited in each silicon nanodisk in the array retrieved through the multipole decomposition are presented. Parameters of the simulated array are the same as in Fig.5 from the main text (disk diameter D = 170 nm, disk height H = 160 nm, array pitch P = 300 nm and substrate refractive index is of 1.45). Figures S10a and b show the radiation patterns of the dipoles at wavelengths of 590 nm and 735 nm leading, respectively, to a minimum reflection at 25 and 60 degrees of incidence, respectively. The radiation patterns were computed with Stratton-Chu equations[5] taking into account the presence of the substrate. For these calculations, a sphere enclosing the dipoles and the substrate was considered. Although the solution is not exact, convergence against variations in the radius of the sphere was checked, yielding almost the same results. Both radiation patterns show minima in the direction of the reflected wave, thus confirming the interference origin of the observed vanishing reflection effect also in the case of silicon disks on substrate.

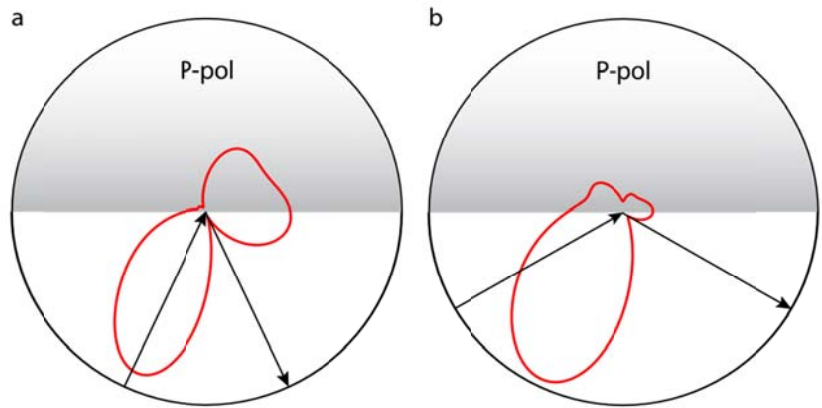

**Figure S10. Radiation patterns in the plane of incidence from electric and magnetic dipoles excited in each silicon nanodisk in the square array of the nanodisks on silica substrate under *p*-polarized incident light. a**, $\lambda = 590$ nm, when reflection vanishes at the incidence angle of 25 degrees. **b**, $\lambda = 735$ nm, when reflection vanishes at the incidence of 60 degrees.

# 9. Explicit expressions used in the Multipole Decomposition.

Multipole decomposition technique was employed to analyze the different modes excited inside the particles. For particles in an array embedded in air, multipoles can be computed through the polarization currents induced inside them:

$$\boldsymbol{J} = -i\omega\varepsilon_0(\varepsilon - 1)\boldsymbol{E},$$

where $\varepsilon$ is the permittivity of the particle and $\boldsymbol{E} = \boldsymbol{E}(\boldsymbol{r})$ the electric field inside it.

This approach fully takes into account mutual interactions in the lattice[6] as well as the possible presence of a substrate. Cartesian basis with origin in the center of the particles was used in the present work. An accurate description of the radiative properties in this basis involves the introduction of the family of toroidal moments[7] and the mean-square radii corrections. Although the explicit expression of the multipoles can be found in some references (see, in particular Ref.7, for the explicit connection with the usual spherical multipole moments) we repeat them here for completeness.

The dipolar moments induced in the system read as:

$$\boldsymbol{p}_{car} = \int \varepsilon_0(\varepsilon - 1)\boldsymbol{E}\,d\boldsymbol{r}$$

$$\boldsymbol{m}_{car} = \frac{-i\omega}{2}\int \varepsilon_0(\varepsilon - 1)[\boldsymbol{r} \times \boldsymbol{E}]\,d\boldsymbol{r}$$

$$\boldsymbol{t} = \frac{-i\omega}{10}\int \varepsilon_0(\varepsilon - 1)[(\boldsymbol{r} \cdot \boldsymbol{E})\boldsymbol{r} - 2r^2\boldsymbol{E}]\,d\boldsymbol{r}$$

and the mean-square radii of the dipole distributions as:

$$\overline{R_m^2} = \frac{-i\omega}{2}\int \varepsilon_0(\varepsilon - 1)[\boldsymbol{r} \times \boldsymbol{E}]r^2\,d\boldsymbol{r}$$

$$\overline{R_t^2} = \frac{-i\omega}{28}\int \varepsilon_0(\varepsilon - 1)[3r^2\boldsymbol{E} - 2(\boldsymbol{r} \cdot \boldsymbol{E})\boldsymbol{r}]r^2\,d\boldsymbol{r}$$

where only the magnetic and toroidal components are considered, since the electric one does not contribute to radiation[7]. For the quadrupolar moments we have the following expressions:

$$\bar{\bar{Q}}_e = \frac{1}{2}\int \varepsilon_0(\varepsilon-1)\left[\mathbf{r}\otimes\mathbf{E} + \mathbf{E}\otimes\mathbf{r} - \frac{2}{3}(\mathbf{r}\cdot\mathbf{E})\bar{\bar{I}}\right]d\mathbf{r}$$

$$\bar{\bar{Q}}_m = \frac{-i\omega}{3}\int \varepsilon_0(\varepsilon-1)[\mathbf{r}\otimes(\mathbf{r}\times\mathbf{E}) + (\mathbf{r}\times\mathbf{E})\otimes\mathbf{r}]d\mathbf{r}$$

$$\bar{\bar{Q}}_t = \frac{-i\omega}{28}\int \varepsilon_0(\varepsilon-1)\left[4(\mathbf{r}\cdot\mathbf{E})\mathbf{r}\otimes\mathbf{r} - 5r^2(\mathbf{r}\otimes\mathbf{E} + \mathbf{E}\otimes\mathbf{r}) + 2r^2(\mathbf{r}\cdot\mathbf{E})\bar{\bar{I}}\right]d\mathbf{r}$$

with $\otimes$ being the dyadic product. It can be shown that both the Cartesian electric dipole and the toroidal dipole have the same radiation pattern. Thus, when using equation (1), the following identifications were made:

$$\mathbf{p} = \mathbf{p}_{car} + \frac{ik_0}{c}\left(\mathbf{t} + \frac{k_0^2}{10}\overline{\mathbf{R}_t^2}\right)$$

$$\mathbf{m} = \mathbf{m}_{car} - k_0^2\overline{\mathbf{R}_m^2}$$

The scattering cross sections in SI units then read:

$$C_{sca}^{(ED)} = \frac{k_0^4}{6\pi\varepsilon_0^2 E_0^2}\left|\mathbf{p}_{car} + \frac{ik_0}{c}\left(\mathbf{t} + \frac{k_0^2}{10}\overline{\mathbf{R}_t^2}\right)\right|^2$$

$$C_{sca}^{(MD)} = \frac{\eta_0^2 k_0^4}{6\pi E_0^2}\left|\mathbf{m}_{car} - k_0^2\overline{\mathbf{R}_m^2}\right|^2$$

$$C_{sca}^{(EQ)} \approx \frac{k_0^6}{80\pi\varepsilon_0^2 E_0^2}\left|\bar{\bar{Q}}_e + \frac{ik_0}{c}\bar{\bar{Q}}_t\right|^2$$

$$C_{sca}^{(MQ)} \approx \frac{\eta_0^2 k_0^6}{80\pi E_0^2}\left|\bar{\bar{Q}}_m\right|^2$$

## 10. Experimental setup scheme.

For completeness we present in Figure S11 a schematic representation of the home-built experimental setup used for measuring the angular transmission/reflection from the fabricated silicon nanodisks arrays. The state of polarization of incident light is described in each of the major steps in the setup. The lamp is used for alignment purposes only and is switch-off during the measurements.

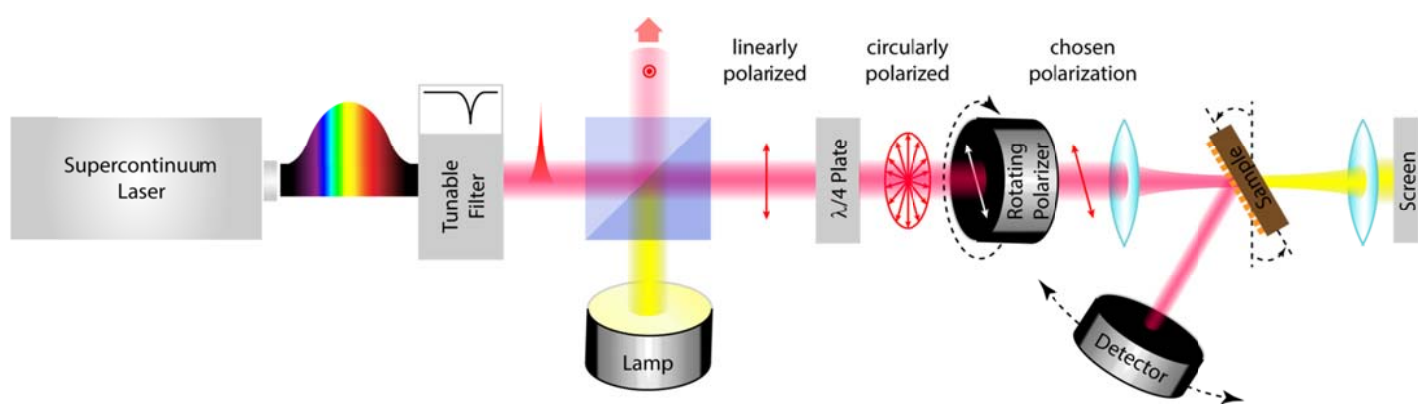

**Figure S11. Scheme of the home-built experimental setup used for measuring angular transmission/reflection from the fabricated nanodisks arrays.**